\documentclass[journal,10pt]{IEEEtran}
\usepackage{amssymb}
\usepackage{amsmath}
\usepackage{cite}
\usepackage{url}
\usepackage{xcolor}
\usepackage{cite,graphicx,amsmath,amssymb}
\usepackage{subfigure}
\usepackage{citesort}
\usepackage{fancyhdr}
\usepackage{mdwmath}
\usepackage{mdwtab}
\usepackage{caption}
\usepackage{amsthm}
\usepackage{setspace}
\usepackage{algorithm}
\usepackage{algorithmic}
\usepackage{makecell}
\usepackage{diagbox}
\usepackage{stfloats}
\usepackage{balance} 
\usepackage{graphicx}
\usepackage{epstopdf}
\usepackage{epsfig}

\newtheorem{remark}{Remark}\newtheorem{theorem}{Theorem}

\newtheorem{lemma}{Lemma}

\newtheorem{corollary}{Corollary}

\allowdisplaybreaks
\makeatletter
\def\ScaleIfNeeded{
	\ifdim\Gin@nat@width>\linewidth \linewidth \else \Gin@nat@width
	\fi } \makeatother
 
\begin{document}
	\title{Pinching-Antenna Systems (PASS)-enabled Secure Wireless Communications}
	\author{
		Guangyu~Zhu,
		Xidong~Mu,
		Li~Guo,
		Shibiao~Xu,
		Yuanwei Liu, \emph{Fellow, IEEE},
		Naofal Al-Dhahir, \emph{Fellow, IEEE}
		\thanks{Guangyu Zhu, Li Guo and Shibiao Xu are with the Key Laboratory of Universal Wireless Communications, Ministry of Education, Beijing University of Posts and Telecommunications, Beijing 100876, China, also with the School of Artificial Intelligence, Beijing University of Posts and Telecommunications, Beijing 100876, China, and also with the National Engineering Research Center for Mobile Internet Security Technology, Beijing University of Posts and Telecommunications, Beijing 100876, China (email:\{Zhugy, guoli, shibiaoxu\}@bupt.edu.cn).}
		\thanks{Xidong Mu is with the Centre for Wireless Innovation (CWI), Queen's University Belfast, Belfast, BT3 9DT, U.K. (e-mail:
			x.mu@qub.ac.uk).}
		\thanks{Yuanwei Liu is with the Department of Electrical and Electronic Engineering, The University of Hong Kong, Hong Kong. (e-mail:
		yuanwei@hku.hk).}
	    \thanks{Naofal Al-Dhahir is with the Department of Electrical and Computer
	    	Engineering, The University of Texas at Dallas, Richardson, TX 75080 USA.
	    	(e-mail: aldhahir@utdallas.edu).}
	}
	\maketitle
	\begin{abstract}
		A novel pinching-antenna systems (PASS)-enabled secure wireless communication framework is proposed. By dynamically adjusting the positions of dielectric particles, namely pinching antennas (PAs), along the waveguides, PASS introduces a novel concept of pinching beamforming to enhance the performance of physical layer security. A fundamental PASS-enabled secure communication system is considered with one legitimate user and one eavesdropper. Both single-waveguide and multiple-waveguide scenarios are studied. 1) For the single-waveguide scenario, the secrecy rate (SR) maximization is formulated to optimize the pinching beamforming. A PA-wise successive tuning (PAST) algorithm is proposed, which ensures constructive signal superposition at the legitimate user while inducing a destructive legitimate signal at the eavesdropper. 2) For the multiple-waveguide scenario, artificial noise (AN) is employed to further improve secrecy performance. A pair of practical transmission architectures are developed: \emph{waveguide division (WD)} and \emph{waveguide multiplexing (WM)}. The key difference lies in whether each waveguide carries a single type of signal or a mixture of signals with baseband beamforming. For the SR maximization problem under the WD case, a two-stage algorithm is developed, where the pinching beamforming is designed with the PAST algorithm and the baseband power allocation among AN and legitimate signals is solved using successive convex approximation (SCA). For the WM case, an alternating optimization algorithm is developed, where the baseband beamforming is optimized with SCA and the pinching beamforming is designed employing particle swarm optimization.
		Numerical results demonstrate that i) PASS can significantly improve the secrecy performance over conventional antenna systems in both scenarios; ii) the proposed PAST algorithm for the single-waveguide scenario is efficient, especially when the number of PAs is even or large; iii) WM provides higher and more stable performance at the cost of increased complexity, while WD serves as a simple yet scalable alternative, which is effective when a large number of PAs are deployed.
	\end{abstract}
\section{Introduction}
With the exponential growth of sensitive data (e.g., personal tracking, financial transactions, and mobile communications) and the rapid expansion of potential attack vectors, information security remains a pressing and long-term challenge for future sixth-generation (6G) networks \cite{6G_security}. While encryption and decryption have long served as fundamental pillars of information security, their high computational complexity and intricate key management increasingly struggle to keep pace with the ever-expanding data demands of the future \cite{cryptography}. To address this issue, physical layer security (PLS) has been proposed as a highly promising technology, garnering significant attention in recent years \cite{Shiu_PLS,Liu_PLS}. Unlike traditional encryption methods, PLS fortifies security at the physical transmission level by exploiting the inherent randomness and uniqueness of wireless channels, such as noise, fading, and interference, to restrict the information available to eavesdroppers \cite{Zou_PLS}. This approach shifts the security burden from algorithmic complexity to physical channel control, enabling lightweight yet robust protection without the need for extensive computational resources. However, its effectiveness is highly dependent on dynamic and environment-sensitive wireless conditions, making precise channel control a key focus for enhancing the PLS performance.

Recent advancements in flexible-antenna technologies, such as reconfigurable intelligent surfaces (RISs) \cite{RIS_survey}, fluid antennas \cite{Wong_Fluid}, and movable antennas \cite{Zhu_MA}, have demonstrated significant potential for enhancing the PLS performance in wireless systems. By dynamically reshaping propagation environments, these technologies improve the legitimate channel quality while suppressing the eavesdropper reception, thereby boosting the secrecy rate (SR) \cite{RIS_PLS,STAR_PLS,Fluid_PLS,MA_PLS}. Nevertheless, these technologies also have their own limitations. For example, RISs suffer from severe double-fading effects due to their passive operation, degrading performance compared to direct line-of-sight (LoS) links \cite{Double_fading}. 
While movable and fluid antennas circumvent this issue, their operation space is limited to just a few wavelengths, restricting their ability to mitigate only the small-scale fading while remaining ineffective against the large-scale path loss. Furthermore, existing flexible-antenna systems have limited reconfigurability, making it challenging to dynamically adjust the number of antennas in terms of practical communication requirements. These constraints pose substantial challenges for the widespread adoption of flexible-antenna technologies in future high-frequency and long-distance communication scenarios, highlighting the need for more adaptive and scalable solutions.
\begin{figure}
	\setlength{\abovecaptionskip}{0cm}   
	\setlength{\belowcaptionskip}{0cm}   
	\setlength{\textfloatsep}{7pt}
	\centering
	\includegraphics[width=3.5in]{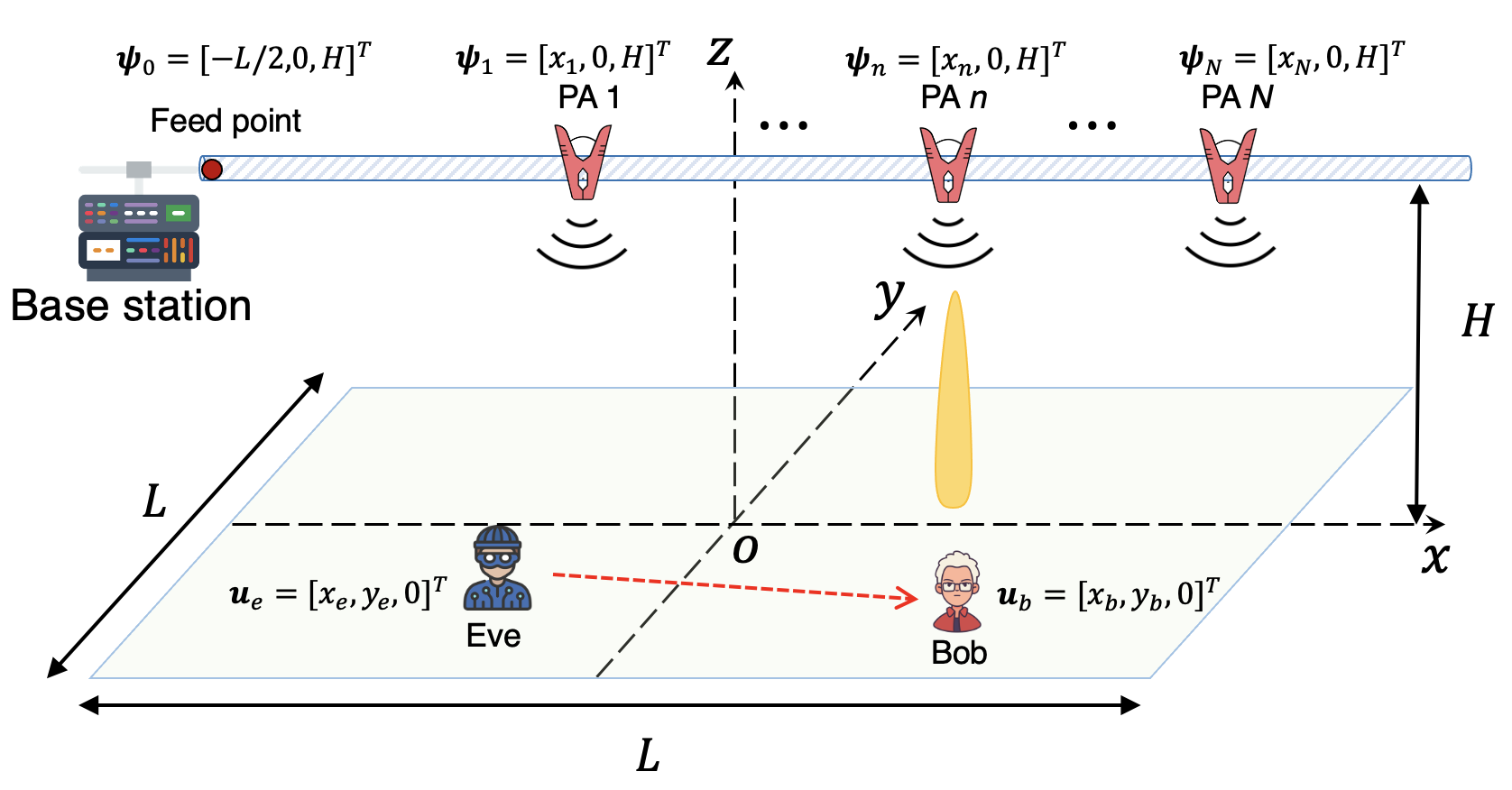}
	\caption{Illustration of a PASS-enabled secure wireless communication system with single-waveguide.}
	\label{SW_system}
\end{figure}

To overcome the above challenges, a novel technology, namely pinching-antenna systems (PASS), has emerged as a promising innovation for flexible-antenna technologies, whose first prototype was demonstrated by NTT DOCOMO \cite{DOCOMO}. As shown in Fig. \ref{SW_system}, PASS employs dielectric waveguides as the primary transmission medium, confining electromagnetic waves to minimize propagation losses over long distances. Along the waveguide,  particles, referred to as pinching antennas (PAs), serve as reconfigurable radiation nodes. When activated via mechanical or electrical adjustments, these PAs can locally perturb the waveguide’s electromagnetic field and couple guided waves into the free space, enabling targeted signal transmission. This design not only supports a flexible LoS link establishment but also significantly reduces the path loss, achieving an efficient spatial control with minimal structural complexity. Furthermore, PASS features a modular, lightweight design that allows PAs to be easily added, removed, or repositioned. This plug-and-play capability enables a rapid reconfiguration of coverage patterns without major hardware changes, making PASS particularly suitable for dynamic or heterogeneous scenarios such as dense urban areas and adaptive indoor networks \cite{Liu_PASS}.

\subsection{Prior Work}
The promising advantages of PASS have begun to attract research interest, leading to initial explorations of its theoretical foundations. Among these studies, the authors of \cite{Ding_PASS} conducted the first theoretical analysis of PASS, exploring the signal transmission model in both single- and multiple-waveguide scenarios and comparing their performance with conventional antenna systems. Consequently, the authors of \cite{zhaolin_PASS} further proposed a physics-based hardware model, where the PA is treated as an open-ended directional coupler, and two power models are derived based on coupled-mode theory. Moreover, the authors of \cite{Tyrovolas_PASS} analytically assessed PASS performance via closed-form expressions for outage probability and average rate. Beyond precise model construction and performance analysis, optimizing PA positions has emerged as another critical research focus in PASS. In \cite{Xu_PASS}, a two-stage optimization framework was developed for downlink PASS, where PA positions were first adjusted to minimize the large-scale path loss, followed by a refinement stage to maximize received signal strength. Additionally, \cite{Tegos_PASS} developed an efficient approach that decouples PA positioning and resource allocation to optimize the minimum uplink data rate in PASS.

\subsection{Motivations and Contributions}
Existing research on PASS primarily focuses on basic communication systems, with limited exploration of its integration into advanced applications such as secure transmission. However, as PLS becomes increasingly critical in modern wireless systems, the unique characteristics of PASS present new opportunities for enhancing secure communications. On the one hand, strategically adjusting the positions of PAs can enlarge the channel disparity between legitimate users and eavesdroppers, thereby improving the SR. On the other hand, with multiple waveguides, joint beamforming between the baseband and PAs can further boost signal quality for legitimate users while effectively suppressing the unauthorized interception. Moreover, unlike conventional antenna systems, PASS offers a higher design flexibility and lower deployment costs, as it supports not only the repositioning of PAs but also the flexible adjustment of waveguide spacing, which in turn enables a broader range of secure communication strategies. Despite these advantages, the application of PASS in PLS remains largely underexplored. Motivated by this gap, we introduce PASS into the PLS framework and summarize our main contributions as follows:
\begin{itemize}
	\item We explore the potential of PASS for enhancing the performance of PLS. A PASS-enabled secure communication system is considered, which consists of a base station (BS) employing PASS, a single-antenna legitimate user (Bob), and a single-antenna eavesdropper (Eve). Both single-waveguide and multiple-waveguide scenarios are studied. For the multiple-waveguide case, artificial noise (AN) is further employed and a pair of practical PASS transmission architectures are developed.
	\item For the single-waveguide scenario, we formulate an SR maximization problem by optimizing the pinching beamforming design. To tackle this non-convex problem, we propose a PA-wise successive tuning (PAST) algorithm. We first determine the optimal positions of the PAs by minimizing the sum of the inverse distances from all PAs to Bob, thereby reducing large-scale path loss. Based on this initial positioning, we then fine-tune the PA positions to ensure constructive signal superposition at Bob while inducing destructive interference at Eve.
	\item For the multiple-waveguide scenario, AN is employed to further improve secrecy performance. We develop a pair of practical transmission architectures, namely waveguide division (WD) and waveguide multiplexing (WM), and formulate the corresponding SR maximization problems. For the WD case, we propose a two-stage algorithm, where the pinching beamforming is designed employing the PAST algorithm and the baseband power allocation among AN and legitimate signals is solved using successive convex approximation (SCA). For the WD case, we develop an alternating optimization (AO) algorithm employing the SCA and particle swarm optimization (PSO) methods to optimize the baseband beamforming and pinching beamforming in an iterative manner.
	\item Our numerical results demonstrate that 1) PASS can achieve a more significant secrecy performance gain than conventional antenna systems; 2) in the single-waveguide scenario, the proposed two-stage algorithm offers high efficiency, particularly when the number of PAs is even or relatively large; and 3) in the multiple-waveguide scenario, WM provides superior secrecy performance via the enhanced joint design, while WD enables a low-complexity and scalable option with competitive performance achieved.
\end{itemize}
\subsection{Organization and Notations}
The remainder of this paper is organized as follows: Section II studies the single-waveguide PASS-enabled PLS, where an SR maximization problem is formulated and a low-complexity PAST algorithm is proposed for the pinching beamforming design. In Section III, the multiple-waveguide PASS-enabled PLS is investigated, where a pair of practical PASS transmission architectures is proposed. For each architecture, the SR is maximized with the corresponding developed algorithm. Section IV presents numerical results and detailed discussions. Finally, Section V concludes the paper.

\emph{Notations}: Scalars, vectors, and matrices are denoted by lower-case, bold lower-case letters, and bold upper-case letters, respectively. $(\cdot)^T$ denotes the transpose, while $(\cdot)^H$ denotes the conjugate transpose. $\mathrm{Tr}(\cdot)$ and $\mathrm{Rank}(\cdot)$ denote the trace and rank of the matrices, respectively. $\|\cdot\|$ and $\|\cdot\|_2$ denote the norm and spectral norm, respectively. $\mathrm{diag}(\cdot)$ denotes the diagonalization operation on vectors. $\mathbb{C}^{M \times N}$ denotes the space of $M \times N$ complex valued matrices. $\mathbf{I}^{M\times M}$ denotes the unit matrix of order $M$. $\mathcal{CN}(\mu, \sigma^2)$ represents the distribution of a circularly symmetrical complex Gaussian random variable with a mean of $\mu$ and a variance of $\sigma^2$. 
\section{Single-Waveguide PASS-enabled Secure Wireless Communications}
This section studies a single-waveguide PASS-enabled communication system, where the pinching beamforming is optimized to boost the SR.
\subsection{System Model}
As illustrated in Fig. \ref{SW_system}, we consider a fundamental downlink secure wireless system consisting of a BS, a single-antenna Bob, and a single-antenna Eve. Specifically, a novel PASS technology is implemented at the BS, where $N$ PAs are strategically deployed across the dielectric waveguide, whose set is denoted by $\mathcal{N}=\{1,\cdots,N\}$. Within a 3D Cartesian coordinate system, Bob and Eve are randomly located in a square region of size $L \times L$ centered at the origin, with positions $\mathbf{u}_b = [x_b,y_b,0]^T$ and $\mathbf{u}_e = [x_e,y_e,0]^T$, respectively. A waveguide of length $L$ is positioned parallel to the $x$-axis at a height of $H$. Accordingly, the location of the $n$-th PA on the waveguide is given by $\boldsymbol{\psi}_{n}=[x_{n},0,H]^T$, where $-L/2 \leq x_{n} \leq L/2$. Without loss of generality, we assume that the PAs are deployed in a successive order, meaning $x_n > x_{n-1}, \forall n \in \mathcal{N}$. 

Given that the PAs are positioned to establish LoS links with users, the free-space channel vectors between the PAs and user $k\in\{b,e\}$ can be expressed as follows
\begin{align}
	\widetilde{\mathbf{h}}_k(\mathbf{x})\!=\bigg[\frac{\eta^{\frac{1}{2}}\!e^{-j\frac{2\pi}{\lambda}\|\mathbf{u}_k-\boldsymbol{\psi}_{1}\|}}{\|\mathbf{u}_k-\boldsymbol{\psi}_{1}\|}\!,\cdots\!,\!\frac{\eta^{\frac{1}{2}}\!e^{-j\frac{2\pi}{\lambda}\|\mathbf{u}_k-\boldsymbol{\psi}_{N}\|}}{\|\mathbf{u}_k-\boldsymbol{\psi}_{N}\|}\bigg]^T.
\end{align}
Here, if $k=b$, it refers to the channel between the PAs and Bob; otherwise, it refers to the channel between the PAs and Eve. Moreover, $\sqrt{\eta}=\frac{c}{4\pi f_c}$ denotes the path loss coefficient where $c$ is the speed of light and $f_c$ is the carrier frequency $f_c$. Similarly, $\lambda=\frac{c}{f_c}$ represents the wavelength corresponding to the carrier frequency. $\mathbf{x}=[x_{1},\cdots,x_{N}]^T$ represents the $x$-axis position vector of all $N$ PAs on the waveguide. $\|\mathbf{u}_k-\boldsymbol{\psi}_{n}\|=\sqrt{(x_k-x_{n})^2+y_k^2+H^2}$ denotes the distance from the $n$-th PA of the waveguide to user $k\in \{b,e\}$. 

In addition, the channel vector within the waveguide, spanning from the feed point to all PAs, is given by
\begin{align}
	\mathbf{e}(\mathbf{x})\!=\!\bigg[e^{-j\frac{2\pi}{\lambda_g}\left\|\boldsymbol{\psi}_{0}-\boldsymbol{\psi}_{1}\right\|},\cdots\!,e^{-j\frac{2\pi}{\lambda_g}\left\|\boldsymbol{\psi}_{0}-\boldsymbol{\psi}_{N}\right\|}\bigg]^T,
\end{align}
where $\boldsymbol{\psi}_{0}$ represents the location of the feed point of the waveguide, whose coordinates are given by $[-L/2,0,H]^T$. Then, $\left\|\boldsymbol{\psi}_{0}-\boldsymbol{\psi}_{n}\right\|=(x_{n}+L/2)$ denotes the distance from the feed point to the $n$-th PA along the waveguide. Besides, we note that  $\lambda_g=\frac{\lambda}{n_{eff}}$, where $n_{eff}$ denotes the effective refractive index of a dielectric waveguide \cite{Ding_PASS}. As a consequence, the complete channel coefficient from the BS to user $k \in \{b,e\}$ can be expressed as follows
\begin{align}
	h_{k}(\mathbf{x})&=\widetilde{\mathbf{h}}^H_k(\mathbf{x})\mathbf{e}(\mathbf{x}) \nonumber \\
	&=\sum^{N}_{n=1}\frac{\eta^{\frac{1}{2}e^{-2\pi j \left(\frac{1}{\lambda}\left\|\mathbf{u}_k-\boldsymbol{\psi}_{n}\right\|+\frac{1}{\lambda_g}\left\|\boldsymbol{\psi}_{0}-\boldsymbol{\psi}_{n}\right\|\right)}}}{\|\mathbf{u}_k-\boldsymbol{\psi}_{n}\|}.
\end{align}
To characterize the maximum performance limits of PASS-enabled secure communications, we assume that the BS has perfect knowledge of all relevant channel state information (CSI) for the design of the pinching beamforming. This assumption is reasonable in scenarios where the eavesdropper is an active but untrusted user in the system \cite{CSI}.

Let $s$ denote the signal emitted from the baseband with $\mathbb{E}\left[|s|^2\right]=1$. The total transmit power at the BS is denoted by $P$, which is assumed to be uniformly distributed among the $N$ PAs \cite{Ding_PASS}. Under this assumption, the signal received by the user $k\in\{b,e\}$ can be expressed as follows
\begin{align}
	y_k=\sqrt{\frac{P}{N}}h_k(\mathbf{x})s+n_k, k\in\{b,e\},
\end{align}
where $n_k \sim \mathcal{CN}(0, \sigma_k^2)$ denotes the additive white Gaussian noise (AWGN) at user $k\in\{b,e\}$. Here, $P\leq P_{\max}$, where $P_{\max}$ denote the maximum transmit power at the BS. Accordingly, the achievable rate (bit/s/Hz) for user $k\in\{b,e\}$ is given by
\begin{align}
	R_k&={\log_2}{\left(1+\frac{\big|\sqrt{\frac{P}{N}}h_k(\mathbf{x})\big|^2}{\sigma^2_k}\right)} \nonumber \\
	&=\log_2\!\!\left(\!1\!+\!\frac{\eta P}{N\sigma_k^2}\left|\sum_{n=1}^{N}\!\frac{e^{-2\pi j\left(\frac{1}{\lambda}\left\|\mathbf{u}_k-\boldsymbol{\psi}_{n}\right\|+\frac{1}{\lambda_g}\left\|\boldsymbol{\psi}_0-\boldsymbol{\psi}_{n}\right\|\right)}}{\|\mathbf{u}_k-\boldsymbol{\psi}_{n}\|}\right|^2\right)\!.	
\end{align}
Furthermore, the SR in considered system is calculated as follows
\begin{align}
	R_S=\left[R_b-R_e\right]^+,
\end{align}
where $\left[\cdot \right ]^+=\max\left(\cdot,0\right)$.

\subsection{Problem Formulation}
In this paper, we aim to maximize the SR by jointly optimizing the transmit power $P$ and the pinching beamforming $\{\mathbf{x}\}$, subject to constraints of maximum power budget at the BS and the minimum spacing between PAs. Then, the optimization problem is formulated as follows
\begin{subequations}\label{Problem_SW}
	\begin{align}
		&\label{P_SW_C0}\  \max_{\mathbf{x},P} \ R_S  \\
		&\label{P_SW_C1} \ \ \ {\rm s.t.}   \quad P \leq P_{\max}, \\
		&\label{P_SW_C2} \quad \quad  \quad -L/2<x_1<\cdots<x_N<L/2, \\
		&\label{P_SW_C3} \quad \quad \ \quad x_n-x_{n-1}\geq \Delta, \forall n \in \mathcal{N},
	\end{align}
\end{subequations}
where $\Delta$ is the minimum spacing between PAs to avoid the coupling effect. Due to the complex dependence of the channel characteristics on the PA positions, the resulting problem is non-convex and difficult to be solved directly.
\begin{remark}
	Compared with conventional antenna systems, PASS can ensure that the channel gain of Bob consistently exceeds that of Eve through proper pinching beamforming design. As a result, the objective function in problem \eqref{Problem_SW} becomes monotonically increasing with respect to $P$. This implies that the power constraint \eqref{P_SW_C1} is always active at the optimum, i.e., $P = P_{\max}$. Therefore, in the subsequent analysis of this section, we focus solely on optimizing the pinching beamforming vector ${\mathbf{x}}$.
\end{remark}
\subsection{Proposed PAST Algorithm}
It can be observed that maximizing the SR in problem \eqref{Problem_SW} is equivalent to maximizing the signal-to-noise ratio (SNR) gap between Bob and Eve. To facilitate this objective, we first express the SNR of each user as follows
\begin{align}
	\label{SNR_b} \textup{SNR}_b=\frac{\eta P}{N\sigma^2_b}\bigg|\sum^N_{n=1}\frac{e^{-j\theta_n}}{\|\mathbf{u}_b-\boldsymbol{\psi}_n\|}\bigg|, \\
	\label{SNR_e} \textup{SNR}_e=\frac{\eta P}{N\sigma^2_e}\bigg|\sum^N_{n=1}\frac{e^{-j\phi_n}}{\|\mathbf{u}_e-\boldsymbol{\psi}_n\|}\bigg|,
\end{align}
where $\theta_n=\frac{2\pi}{\lambda}\|\mathbf{u}_b-\boldsymbol{\psi}_n\|+\frac{2\pi}{\lambda_g}\|\boldsymbol{\psi}_0-\boldsymbol{\psi}_n\|$ and $\phi_n=\frac{2\pi}{\lambda}\|\mathbf{u}_e-\boldsymbol{\psi}_n\|+\frac{2\pi}{\lambda_g}\|\boldsymbol{\psi}_0-\boldsymbol{\psi}_n\|$ denote the phase-shifts due to wave propagation both inside the waveguide and in the free space. Given the optimal transmit power, the problem further reduces to maximizing the channel gain difference between Bob and Eve, which is jointly determined by large-scale path loss and the phase alignment among the PAs. 

\begin{figure}[t]
 	\subfigure[$\theta_n$.]{\label{Theta}
 		\includegraphics[width= 3.4in]{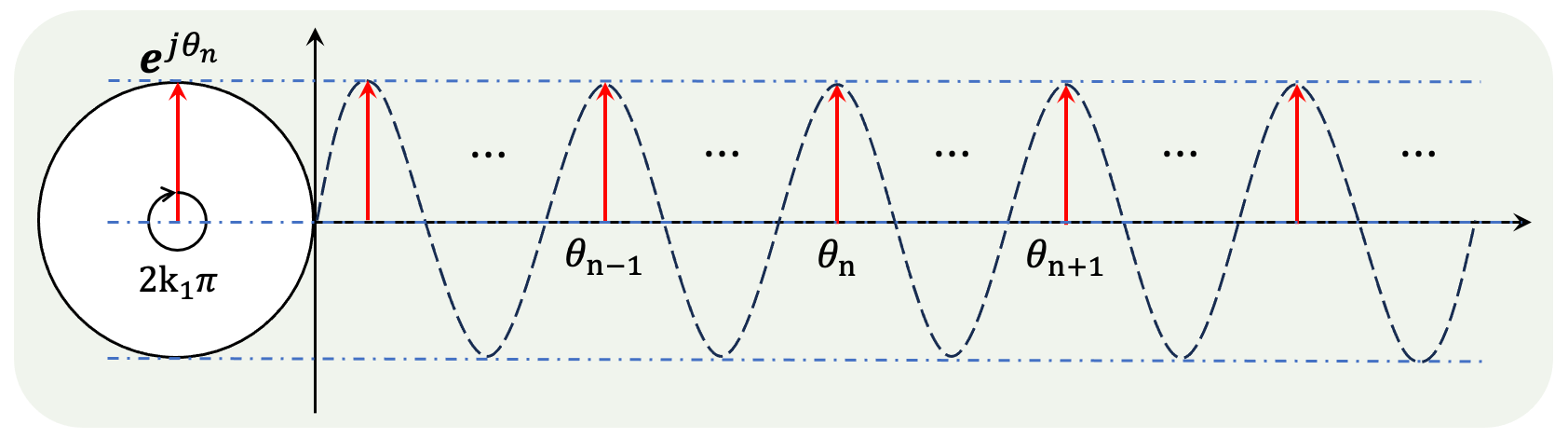}}
 	\subfigure[$\phi_n$.]{\label{Phi}
 		\includegraphics[width= 3.4in]{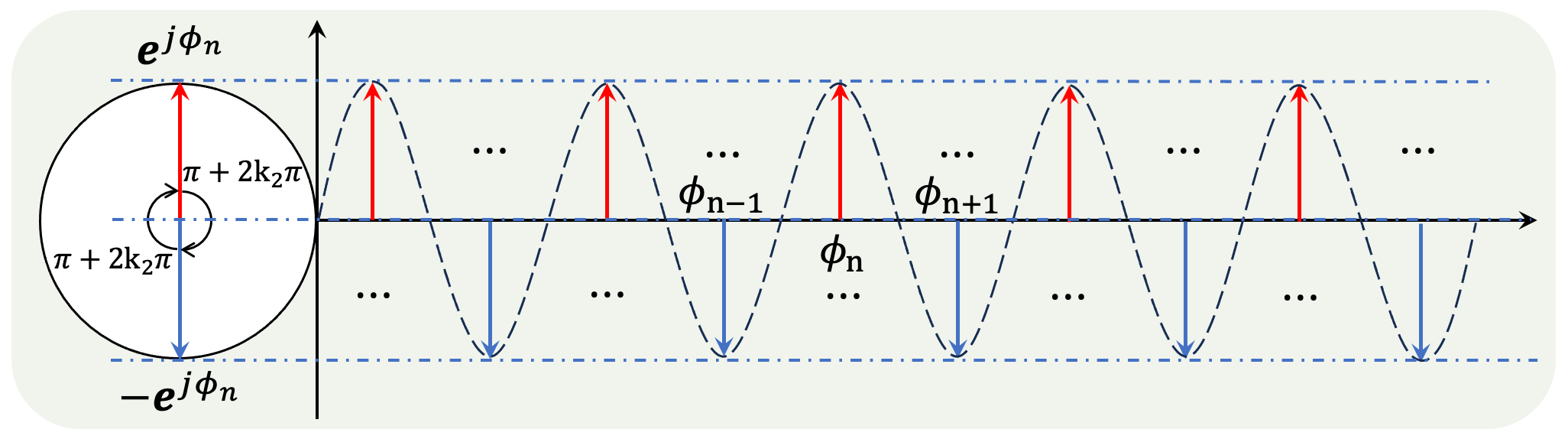}}
 	\caption{\textcolor{black}{Phase alignment strategy.}}
 	\label{Phase}
 \end{figure}

Notably, in mmWave and THz bands, both the free-space wavelength $\lambda$ and the waveguide wavelength $\lambda_g$ are typically on the order of sub-centimeters or smaller, which are significantly shorter than the distance between PAs and users. As a result, even slight PA position adjustments can induce a periodic variation in $\theta_n$ and $\phi_n$. Exploiting this phase sensitivity, we can design a flexible phase alignment strategy to enhance the secrecy performance, as illustrated in Fig. \ref{Phase}:
\begin{itemize}
	\item \textbf{Legitimate Signal Enhancement at Bob:} As shown in Fig. \ref{Theta}, by precisely tuning the PA positions, the phase terms $\theta_n$ can be coherently aligned, enabling constructive combining of the legitimate signal in \eqref{SNR_b}, and thus significantly boosting the received power at Bob. Mathematically, this constructive alignment corresponds to the following constraint
	\begin{align}\label{P_SW_R_C1} 
		\theta_n-\theta_{n-1}=2k_1\pi, k_1 \in \mathcal{Z}, n \in \mathcal{N},
	\end{align}
    where $k_1$ is an arbitrary integer ensuring that adjacent phase shifts differ by integer multiples of $2\pi$.
	\item \textbf{Legitimate Signal Mitigation at Eve:} As shown in Fig. \ref{Phi}, the phase terms $\phi_n$ can be configured in a pairwise-opposite manner to induce destructive combining at Eve. For even numbers of PAs, this yields almost complete signal cancellation in \eqref{SNR_e}; for odd numbers, only a single residual term remains, effectively minimizing information leakage. Additionally, the leaked signal power diminishes as $N$ increases. This destructive alignment requires the phase terms to satisfy
	\begin{align}\label{P_SW_R_C2} 
		\phi_n-\phi_{n-1}=\pi+2k_2\pi, k_2 \in \mathcal{Z}, n \in \mathcal{N},
	\end{align}
 where $k_2$ is an arbitrary integer ensuring anti-phase alignment.
\end{itemize}
This decoupling advantage allows us to focus the large-scale attenuation optimization solely on minimizing Bob’s path loss, without considering Eve during this stage. Based on this insight, the original problem can be relaxed and reformulated as follows
\begin{subequations}\label{Problem_SW_relaxed}
	\begin{align}
		&\label{P_SW_R_C0}\  \max_{\{\mathbf{x}\}}\quad \sum^N_{n=1}\frac{1}{\|\mathbf{u}_b-\boldsymbol{\psi}_n\|}\\ 
	    &\label{P_SW_R_C3} \quad {\rm s.t.}\ \ \eqref{P_SW_C2}, \eqref{P_SW_C3}, \eqref{P_SW_R_C1}, \eqref{P_SW_R_C2}.
	\end{align}
\end{subequations}
Here, the objective function \eqref{P_SW_R_C0} aims to mitigate large-scale path loss by maximizing the sum of the inverse distances from all PAs to Bob. Since PA position adjustments impact the objective function and phase-shift constraints at different scales, we propose a PAST algorithm to efficiently tackle the problem. The first step determines the overall PA distribution by maximizing the objective function without considering \eqref{P_SW_R_C1} and \eqref{P_SW_R_C2}, followed by successively fine-tuning the PA position to satisfy both the constructive and destructive phase alignment constraints.

\emph{1) Coarse PA Distribution for Bob’s Large-Scale Path Loss Minimization:} By neglecting constraints \eqref{P_SW_R_C1} and \eqref{P_SW_R_C2}, problem \eqref{Problem_SW_relaxed} simplifies to
\begin{subequations}\label{Problem_SW_PL}
	\begin{align}
			&\label{P_SW_PL_C0}\  \max_{\{\mathbf{x}\}}\quad \sum^N_{n=1}\frac{1}{\sqrt{(x_n-x_b)^2+D^2_b}}\\ 
			&\label{P_SW_PL_C1}\quad {\rm s.t.} \quad \eqref{P_SW_R_C1}, \eqref{P_SW_R_C2}.
	\end{align}
\end{subequations}
where $D_b=\sqrt{y^2_b+H^2}$. To solve this problem, the following lemma provides valuable insights.
\begin{lemma}
	When $D_b \gg \Delta$, the optimal solution of $\mathbf{x}$ is a uniform symmetric arrangement of the $N$ PAs, centered at $x_m$ and spaced at $\Delta$, i.e., 
	\begin{align}
		x^*_n=x_m-\frac{(N-1)}{2}\Delta+(i-1)\Delta, i=1,\cdots,N.
	\end{align}
\textbf{Proof:} See Appendix A.
\end{lemma}
\emph{2) Successive Fine-Tuning for Phase-Constrained Beamforming:} With the coarse positions of all PAs determined, the remaining task is to refine their locations to satisfy the phase alignment constraint in \eqref{P_SW_R_C3}. For analytical tractability, we express $\theta_n$ and $\phi_n$ in the following specific form
\begin{align}
	\label{theta} \theta_n=\frac{2\pi}{\lambda}\sqrt{(x_n-x_b)^2+D^2_b}+\frac{2\pi}{\lambda_g}(x_n+L/2),\\ 
	\label{phi} \phi_n=\frac{2\pi}{\lambda}\sqrt{(x_n-x_e)^2+D^2_e}+\frac{2\pi}{\lambda_g}(x_n+L/2), 
\end{align}
where $D_e=\sqrt{y^2_e+H^2}$. To ensure the accuracy of the phase approximation and avoid significant deviation from the optimized path loss, only slight adjustments are made to the positions of neighboring PAs. As a result, we can approximate $\theta_n$ using the first-order Taylor expansion around $\theta_{n-1}$ as follows
\begin{align}
	\theta_n &\approx \frac{2\pi}{\lambda}\frac{x_{n-1}-x_b}{\sqrt{(x_{n-1}-x_b)^2+D^2_b}}(x_n-x_{n-1}) \nonumber \\
	&+\frac{2\pi}{\lambda}\sqrt{(x_{n-1}-x_b)^2+D^2_b}+\frac{2\pi}{\lambda_g}(x_{n}+L/2),
\end{align}
In this case, constraint \eqref{P_SW_R_C1} can be further rewritten as
\begin{align}\label{k1}
	\Delta x\bigg[\frac{1}{\lambda}\frac{x_{n-1}-x_b}{\sqrt{(x_{n-1}-x_b)^2+D^2_b}}+\frac{1}{\lambda_g}\bigg]=k_1,
\end{align}
where $\Delta x=x_n-x_{n-1}$. Similarly, constraint \eqref{P_SW_R_C2} can be converted to
\begin{align}\label{k2}
	\Delta x\bigg[\frac{1}{\lambda}\frac{x_{n-1}-x_e}{\sqrt{(x_{n-1}-x_e)^2+D^2_e}}+\frac{1}{\lambda_g}\bigg]=\frac{1}{2}+k_2.
\end{align}
By combining the two equations, we obtain the following expression as the basis for fine-tuning
\begin{align}
	\widehat{\Delta} x=\lambda\frac{k_1-1/2-k_2}{\frac{x_{n-1}-x_b}{\sqrt{(x_{n-1}-x_b)^2+D^2_b}}-\frac{x_{n-1}-x_e}{\sqrt{(x_{n-1}-x_e)^2+D^2_e}}}.
\end{align}
Since all terms involving $\lambda_g$ are canceled out during the combination of the two expressions, the resulting $\widehat{\Delta} x$ does not fully satisfy the original constraints \eqref{k1} and \eqref{k2}. Therefore, an additional refinement of $\widehat{\Delta} x$ is necessary to restore feasibility. Next, substituting $\widehat{\Delta} x$ into constraint (18) yield
\begin{align}
	\widehat{\Delta} x\bigg[\frac{1}{\lambda}\frac{x_{n-1}-x_b}{\sqrt{(x_{n-1}-x_b)^2+D^2_b}}+\frac{1}{\lambda_g}\bigg]=k_1+\Delta {k_1},
\end{align}
where $\Delta k_1$ denotes a deviation term that needs to be compensated. Given that the first term on the left-hand side is negligible compared to the second, we concentrate on adjusting the latter, leading to the following refined expression
\begin{align}
	\widetilde{\Delta} x = -\lambda_g\Delta k_1.
\end{align}
Accordingly, the expression for $\Delta x$ satisfying the constraint \eqref{k1} is given by
\begin{align} \label{Delta1}
	\Delta x = \widehat{\Delta} x + \widetilde{\Delta} x.
\end{align}
In addition, the impact of $\widetilde{\Delta} x$ on the first term on the left-hand side of the equation \eqref{k2} is negligible, allowing the constraint to be considered satisfied with $\Delta x$. Therefore, given the position of the $(n-1)$-th PA $x_{n-1}$, we can refine the $n$-th PA’s position as
\begin{align}\label{x_R}
	x_n=x_{n-1}+\Delta x.
\end{align}
Conversely, for a given $x_n$, we can refine $x_{n-1}$ in a similar way as follows:
\begin{gather} 
		\label{x_L} x_{n-1}=x_{n}-\Delta' x,\\
		\label{Delta2} \Delta' x=\lambda\frac{k_1-1/2-k_2}{\frac{x_{n}-x_b}{\sqrt{(x_{n}-x_b)^2+D^2_b}}-\frac{x_{n}-x_e}{\sqrt{(x_{n}-x_e)^2+D^2_e}}}-\lambda_g\Delta k_1.
\end{gather}
Note that, to minimize the disturbance to the large-scale path loss determined in the first stage, we select the PA closest to $x_b$, i.e., $x_{\lfloor \frac{N+1}{2}\rfloor}$, as the reference point for refining the positions of the remaining PAs.  Meanwhile, the step sizes $\Delta x$ and $\Delta' x$ are chosen as the minimum values among all feasible solutions satisfying $\Delta x \geq \Delta$ and $\Delta' x \geq \Delta$. 

To this end, the overall algorithm for solving problem \eqref{Problem_SW_relaxed} is summarized in \textbf{Algorithm \ref{alg:A}}. By leveraging the first-order Taylor expansion and the proposed PAST, the overall computational complexity for this algorithm is $\mathcal{O}(N)$.

\begin{algorithm}[!t]\label{Algorithm 1}
	\caption{Proposed PAST Algorithm for Solving Problem \eqref{Problem_SW_relaxed}.}
	\label{alg:A}
	\begin{algorithmic}[1]
		\STATE {Solve \eqref{Problem_SW_PL} to determine the reference PA position $x_{\lfloor \frac{N+1}{2} \rfloor}$, and initiate successive fine-tuning from this point.} \\
		\STATE Denote $n_R=x_{\lfloor \frac{N+1}{2}\rfloor}+1$ and $n_L=x_{\lfloor \frac{N+1}{2}\rfloor}-1$.
		\REPEAT
		\STATE Update $x_{n_R}$ using \eqref{Delta1} and \eqref{x_R}; 
		\STATE Update $x_{n_L}$ using \eqref{x_L} and \eqref{Delta2}; 
		\STATE Update $n_R \leftarrow n_R+1$ and $n_L \leftarrow n_L-1$.\\
		\UNTIL $n_R > N$ and $n_L < 1$.
		\STATE {Output} the optimal solutions $\mathbf{x}^*$.
	\end{algorithmic}
\end{algorithm}

\section{Multiple-Waveguide PASS-enabled Secure Wireless Communications}
In this section, we extend our study to a multiple-waveguide PASS-enabled secure wireless communication system. In this scenario, AN can be introduced at the baseband to further enhance secrecy performance \cite{Pigi_PASS}. To support this enhancement, we propose two practical transmission architectures, namely WD and WM, which enable the joint optimization of baseband signal processing and pinching beamforming. Since we continue to focus on a fundamental system model with a single legitimate user and one eavesdropper, a dual-waveguide configuration is adopted as a representative case to gain key design insights in the following.

Here, we adopt the same user distribution as in the single-waveguide scenario, but extend the system to a dual-waveguide configuration. Specifically, each waveguide is equipped with $N$ dielectric particles to activate the PAs, with their index set denoted by $\mathcal{N}$. Both waveguides have identical length $L$ and are placed parallel to the $x$-axis within a horizontal plane at a height of $H$. Accordingly, the location of the $n$-th PA on the $m$-th waveguide is denoted by $\boldsymbol{\psi}_{m,n}=[x_{m,n},y_{m},H]^T$, where $-L/2 \leq x_{m,n} \leq L/2$, $m\in\{1,2\}, n\in \mathcal{N}$.

Similarly, the end-to-end transmission from the baseband to user $k\in\{b,e\}$ via the $m$-th waveguide can be decomposed into two components: in-waveguide propagation from the feed point to the PA, and free-space propagation from the PA to the user. Accordingly, the channel from the $m$-th waveguide to user $k\in\{b,e\}$ can be expressed as follows
\begin{align}
	h_{k}(\mathbf{x}_m)&=\widetilde{\mathbf{h}}^H_k(\mathbf{x}_m)\mathbf{e}(\mathbf{x}_m) \nonumber \\
	&=\sum^{N}_{n=1}\frac{\eta^{\frac{1}{2}e^{-2\pi j \left(\frac{1}{\lambda}\left\|\mathbf{u}_k-\boldsymbol{\psi}_{m,n}\right\|+\frac{1}{\lambda_g}\left\|\boldsymbol{\psi}_{m,0}-\boldsymbol{\psi}_{m,n}\right\|\right)}}}{\|\mathbf{u}_k-\boldsymbol{\psi}_{m,n}\|}.
\end{align}
where $\mathbf{x}_m=[x_{m,1},\cdots,x_{m,N}]^T$ represents the $x$-axis position vector of all $N$ PAs on the $m$-th waveguide. $\|\mathbf{u}_k-\boldsymbol{\psi}_{m,n}\|=\sqrt{(x_k-x_{m,n})^2+(y_k-y_m)^2+H^2}$ denotes the distance from the $n$-th PA of the $m$-th waveguide to user $k\in\{b,e\}$. Besides, $\boldsymbol{\psi}_{m,0}$ denotes the location of the feed point of the $m$-th waveguide, whose coordinates are denoted by $[-L/2,y_m,H]^T$. Then, $\left\|\boldsymbol{\psi}_{m,0}-\boldsymbol{\psi}_{m,n}\right\|=(x_{m,n}+L/2)$ denotes the distance from the feed point to the $n$-th PA of the $m$-th waveguide.

Hence, the end-to-end channel from the BS to user $k\in\{b,e\}$ can be modeled as follows
\begin{align}
	\mathbf{h}_k=[h_k(\mathbf{x}_1),h_k(\mathbf{x}_2)]^T.
\end{align}

\subsection{Practical Transmission Architecture}
In this subsection, we introduce two practical transmission architectures, namely WD and WM, along with their corresponding signal transmission models.
\subsection{System Model}
\begin{figure}[t]
	\subfigure[Waveguide division (WD).]{\label{WD}
		\includegraphics[width= 3.3in]{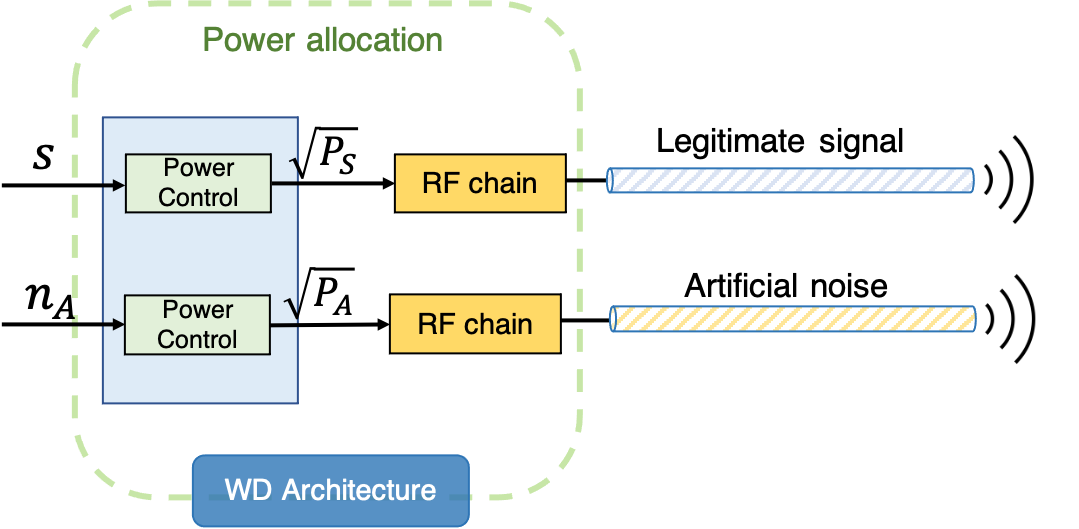}}
	\subfigure[Waveguide multiplexing (WM).]{\label{WM}
		\includegraphics[width= 3.3in]{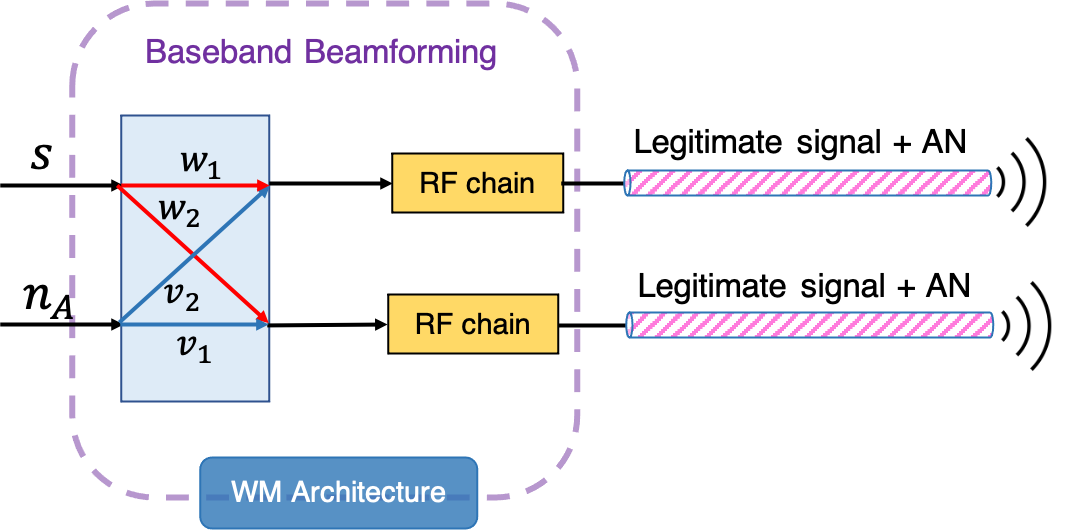}}
	\caption{\textcolor{black}{Illustration of two practical secure transmission architectures.}}
	\label{MW_system}
\end{figure}

\emph{1) Waveguide Division:} As shown in Fig. \ref{WD}, this architecture employs baseband processing capabilities to implement power allocation of two data streams, namely legitimate signal (denoted as $s$) and AN (denoted as $n_A$), and then divisionally feeds them to different waveguides for transmission. Since baseband only performs simple power management without multiplexing, each waveguide exclusively carries a single data stream. To simplify the subsequent analysis, we assume that the legitimate signal $s$ is transmitted on waveguide 1, while the AN $n_A$ is transmitted on waveguide 2. To this end, the received signal at user $k\in\{b,e\}$ can be expressed as
\begin{align}
	y_k\!=\!\sqrt{\frac{P_S}{N}}h_k(\mathbf{x}_1)s\!+\!\sqrt{\frac{P_A}{N}}h_k(\mathbf{x}_2)n_A\!+\!n_k, k\in\{b,e\}. 
\end{align}
where $P_A$ and $P_S$ denote the power allocated to the legitimate signal and AN, respectively. Subject to the constraint of the maximum transmit power of the BS, we have
\begin{align}\label{power_WD}
	P_A+P_S\leq P_{\max}.
\end{align}
Then, the achievable rate (bit/s/Hz) for user $k\in\{b,e\}$ is given by \eqref{Rate_WD}, as shown at the top of the next page.
\begin{figure*}[t]
	\normalsize
		\setlength{\belowdisplayskip}{-10.3pt}
	\begin{align}\label{Rate_WD}		
		R^{\textup{WD}}_k=\log_2\left(1+\frac{\left|\sqrt{\frac{P_S}{N}}h_k(\mathbf{x}_1)\right|^2}{\left|\sqrt{\frac{P_A}{N}}h_k(\mathbf{x}_2)\right|^2+\sigma^2_k}\right)
		=\log_2\left(1+\frac{\!\frac{\eta P_S}{N}\left|\sum_{n=1}^{N}\!\frac{e^{-2\pi j\left(\frac{1}{\lambda}\left\|\mathbf{u}_k-\boldsymbol{\psi}_{1,n}\right\|+\frac{1}{\lambda_g}\left\|\boldsymbol{\psi}_{1,0}-\boldsymbol{\psi}_{1,n}\right\|\right)}}{\|\mathbf{u}_k-\boldsymbol{\psi}_{1,n}\|}\right|^2}{\!\frac{\eta P_A}{N}\left|\sum_{n=1}^{N}\!\frac{e^{-2\pi j\left(\frac{1}{\lambda}\left\|\mathbf{u}_k-\boldsymbol{\psi}_{2,n}\right\|+\frac{1}{\lambda_g}\left\|\boldsymbol{\psi}_{2,0}-\boldsymbol{\psi}_{2,n}\right\|\right)}}{\|\mathbf{u}_k-\boldsymbol{\psi}_{2,n}\|}\right|^2+\sigma^2_k}\right).
	\end{align}
\end{figure*}

\emph{2) Waveguide Multiplexing:} As shown in Fig. \ref{WM}, this architecture multiplexes the legitimate signal and AN before feeding them into the waveguides, thus facilitating the baseband beamforming to enhance the signal transmission flexibility. Let $w_{m}$ and $v_{m}$ denote the beamforming coefficient assigned to the legitimate signal and AN of the $m$-th waveguide, respectively. As a result, the received signal at user $k\in\{b,e\}$ can be expressed as 
\begin{align}
	y_k=\mathbf{h}^H_k(\mathbf{w}s+\mathbf{v}n_A)+n_k, k\in\{b,e\},
\end{align}
where $\mathbf{w}=[w_1,w_2]^T$ and $\mathbf{v}=[v_1,v_2]^T$. It should be noted that due to the transmit power limitations at the BS, the following constraints must be satisfied
\begin{align}\label{power_WM}
	\|\mathbf{w}\|^2+\|\mathbf{v}\|^2 \leq P_{\max}.
\end{align}
Here, $P_{\max}$ is the maximum transmit power at the BS. In this case, the achievable rate (bit/s/Hz) of user $k\in\{b,e\}$ is given by \eqref{Rate_WM}, as shown at the top of the next page.
\begin{figure*}[t]
	\normalsize
		\setlength{\belowdisplayskip}{-0.3pt}
\begin{align}\label{Rate_WM}
	R^{\textup{WM}}_k=\log_2\left(1+\frac{\left|\mathbf{h}^H_k\mathbf{w}\right|^2}{\left|\mathbf{h}^H_k\mathbf{v}\right|^2+\sigma^2_k}\right)
	=\log_2\left(1+\frac{\!\frac{\eta}{N}\left|\sum_{m=1}^{2}\sum_{n=1}^{N}\!\frac{e^{-2\pi j\left(\frac{1}{\lambda}\left\|\mathbf{u}_k-\boldsymbol{\psi}_{m,n}\right\|+\frac{1}{\lambda_g}\left\|\boldsymbol{\psi}_{m,0}-\boldsymbol{\psi}_{m,n}\right\|\right)}}{\|\mathbf{u}_k-\boldsymbol{\psi}_{m,n}\|}w_m\right|^2}{\!\frac{\eta}{N}\left|\sum_{m=1}^{2}\sum_{n=1}^{N}\!\frac{e^{-2\pi j\left(\frac{1}{\lambda}\left\|\mathbf{u}_k-\boldsymbol{\psi}_{m,n}\right\|+\frac{1}{\lambda_g}\left\|\boldsymbol{\psi}_{m,0}-\boldsymbol{\psi}_{m,n}\right\|\right)}}{\|\mathbf{u}_k-\boldsymbol{\psi}_{m,n}\|}v_m\right|^2+\sigma^2_k}\right).
\end{align}
\hrulefill\vspace*{0pt}
\end{figure*}

To this end, the SR for both transmission architectures can be given by
\begin{align}\label{SR}
	R^X_S=[R^X_b-R^X_e]^+, X\in\{\textup{WD},\textup{WM}\}.
\end{align}
\subsection{Problem Formulation}
In this section, we again aim to maximize the SR by jointly optimizing baseband power allocation/beamforming and pinching beamforming, subject to constraints of maximum transmit power at the BS and the minimum spacing between PAs for both architectures. Specifically, in the WD architecture, we optimize the power allocation coefficients $P_A$ and $P_S$ along with the pinching beamforming vectors $\{\mathbf{x}_m\}$. In the WM architecture, we optimize the baseband beamforming vectors $\mathbf{w}$ and $\mathbf{v}$ instead of power allocation coefficients, while still optimizing the pinching beamforming vectors $\{\mathbf{x}_m\}$. Then, the optimization problem is formulated as follows
\begin{subequations}\label{Problem_DW}
	\begin{align}
		&\label{P_DW_C0}\ \max_{P_A,P_S,\{\mathbf{x}_1,\mathbf{x}_2\}} R^{\textup{WD}}_S\bigg/\!\! \max_{\mathbf{w},\mathbf{v},\{\mathbf{x}_1,\mathbf{x}_2\}} R^{\textup{WM}}_S\\ 
		&\label{P_DW_C1} {\rm s.t.}  \ P^X \leq P_{\max}, X \in\{\textup{WD},\textup{WM}\},\\
		&\label{P_DW_C2} \ \quad -L/2 \leq x_{m,n} \leq L/2,n\in\mathcal{N},m \in \{1,2\},\\
		&\label{P_DW_C3} \ \quad x_{m,n}-x_{m,n-1} \geq \Delta, n \in \mathcal{N}, m \in\!\{1,2\},
	\end{align}
\end{subequations}
where $P^{\textup{WD}}$ and $P^{\textup{WM}}$ denote the BS transmit power of the WD and WM architectures, shown in the \eqref{power_WD} and \eqref{power_WM}, respectively.

Compared to the single-waveguide scenario, the multi-waveguide scenario introduces two major challenges. First, additional design variables from baseband signal processing must be jointly optimized along with pinching beamforming, resulting in a tightly coupled and more complex optimization problem. Second, the positions of PAs across different waveguides need to be jointly designed, further increasing the system design complexity. Consequently, these factors lead to a highly challenging problem \eqref{Problem_DW}, which existing methods fail to address effectively. To tackle this challenge, we propose efficient algorithms in the following subsections.

\subsection{Proposed Two-Stage Algorithm for the WD Architecture}
In this subsection, we propose a two-stage algorithm for solving the problem under the WD architecture. In the first stage, we determine the optimal pinching beamforming following the method proposed for the single-waveguide scenario. Then, we optimize the baseband power allocation using the SCA method in the second stage.

\emph{1) Pinching Beamforming Design:} Similar to the single-waveguide scenario, each waveguide in the WD architecture carries a single signal stream. Accordingly, we adopt the PAST algorithm to determine the pinching beamforming. The key difference in the WD architecture lies in the introduction of AN, which is transmitted through a dedicated waveguide. This decoupling enables the independent determination of the PA positions on the two waveguides, which can be obtained by solving the following two subproblems
\begin{subequations}\label{Problem_NCT_relaxed1}
	\begin{align}
		&\label{P_NCT_R1_C0}\  \max_{\mathbf{x}_1}\quad \sum^N_{n=1}\frac{1}{\|\mathbf{u}_b-\boldsymbol{\psi}_{1,n}\|}\\ 
		&\label{P_NCT_R1_C1}  {\rm s.t.} \quad  \theta_{1,n}-\theta_{1,n-1}=2k_3 \pi, k_3 \in \mathcal{Z}, n \in \mathcal{N},\\
		&\label{P_NCT_R1_C2}\quad \quad \ \phi_{1,n}-\phi_{1,n-1}=\pi+2k_4\pi, k_4 \in \mathcal{Z}, n \in \mathcal{N},\\
		&\label{P_NCT_R1_C3} \quad \quad \ \eqref{P_DW_C2}, \eqref{P_DW_C3}.
	\end{align}
\end{subequations}
\vspace{-0.5cm}
\begin{subequations}\label{Problem_NCT_relaxed2}
	\begin{align}
		&\label{P_NCT_R2_C0}\  \max_{\mathbf{x}_2}\quad \sum^N_{n=1}\frac{1}{\|\mathbf{u}_b-\boldsymbol{\psi}_{2,n}\|}\\ 
		&\label{P_NCT_R2_C1}  {\rm s.t.} \quad \theta_{2,n}-\theta_{2,n-1}=2k_5\pi, k_5 \in \mathcal{Z}, n \in \mathcal{N},\\
		&\label{P_NCT_R2_C2}  \quad \quad \ \phi_{2,n}-\phi_{2,n-1}=\pi+2k_6\pi, k_6 \in \mathcal{Z}, n \in \mathcal{N},\\
		&\label{P_NCT_R2_C4} \quad \quad \ \eqref{P_DW_C2}, \eqref{P_DW_C3}.
	\end{align}
\end{subequations}
Here, $\theta_{1,n}$ and $\phi_{1,n}$ denote the transmission phase-shifts from the $n$-th PA on waveguide 1 to Bob and Eve, respectively, while $\theta_{2,n}$ and $\phi_{2,n}$ represent the transmission phase-shifts from the $n$-th PA on waveguide 2 to Eve and Bob, respectively. Moreover, $k_3$, $k_4$, $k_5$, and $k_6$ are arbitrary integers. Following the steps employed in \textbf{Algorithm \ref{alg:A}}, we can obtain the solutions of PAs' positions $\{\mathbf{x}_1, \mathbf{x}_2\}$. 

\emph{2) Baseband Power Allocation Optimization:} Given the pinching beamforming, problem is reduced into a power allocation problem as follows
\begin{subequations}\label{Problem_NCT_Power}
	\begin{align}
		&\label{P_NCT_P_C0}\  \max_{P_A,P_S}  R^{\textup{WD}}_b-R^{\textup{WD}}_e\\ 
		&\label{P_NCT_P_C1}\quad \  {\rm s.t.}\ P_A+P_S \leq P_{\max},
	\end{align}
\end{subequations}
It can be found that the objective function is non-convex with respect to $P_A$ and $P_S$, rendering the problem intractable for direct optimization. To overcome this challenge, we adopt the SCA technique to transform the problem into a more tractable form By applying the first-order Taylor expansion, a convex lower bound of the objective function can be derived with \eqref{R_NCT_low}, as shown at the top of the next page, 
\begin{figure*}[t]
	\normalsize
	\setlength{\belowdisplayskip}{-0.3pt}
	\begin{align}\label{R_NCT_low}
		&\quad\log_2\left(1+\!\frac{P_A|h_b(\mathbf{x}_1)|^2}{P_S|h_b(\mathbf{x}_2)|^2+\sigma_b^2}\right)-\log_2\left(1+\frac{P_A|h_e(\mathbf{x}_1)|^2}{P_S|h_e(\mathbf{x}_2)|^2\!+\!\sigma_e^2}\right) \nonumber \\
		&=\!\log_2\!\left(\!P_A|h_b\!(\mathbf{x}_1)|^2\!\!+\!\!P_S|h_b\!(\mathbf{x}_2)|^2\!\!+\!\sigma^2_b\right)\!+\!\log_2\!\left(\!P_S|h_e\!(\mathbf{x}_2)|^2\!\!+\!\sigma^2_e\right)\!\!-\!\log_2\!\left(\!P_S|h_b\!(\mathbf{x}_2)|^2\!\!+\!\sigma^2_b\right)\!\!-\!\log_2\!\left(\!P_A|h_e\!(\mathbf{x}_1)|^2\!\!+\!\!P_S|h_e\!(\mathbf{x}_2)|^2\!\!+\!\sigma^2_e\right) \nonumber \\
		&\geq \log_2\left(P_w|h_b(\mathbf{x}_1)|^2+P_v|h_b(\mathbf{x}_2)|^2+\sigma^2_b\right)+\log_2\left(P_v|h_e(\mathbf{x}_2)|^2+\sigma^2_e\right)-\log_2\left(P^{(j)}_w|h_e(\mathbf{x}_1)|^2+P^{(j)}_v|h_e(\mathbf{x}_2)|^2+\sigma^2_e\right) \nonumber \\
		&-\log_2\left(P^{(j)}_v|h_b(\mathbf{x}_2)|^2+\sigma^2_b\right)\!-\!\frac{|h_b(\mathbf{x}_2)|^2\left(P_v-P^{(j)}_v\right)}{\left(P^{(j)}_v|h_b(\mathbf{x}_2)|^2+\sigma^2_b\right)\ln 2}\!-\!\frac{|h_e(\mathbf{x}_1)|^2\left(P_w-P^{(j)}_w\right)+|h_e(\mathbf{x}_2)|^2\left(P_v-P^{(j)}_v\right)}{\left(P^{(j)}_w|h_e(\mathbf{x}_1)|^2+P^{(j)}_v|h_e(\mathbf{x}_2)|^2+\sigma^2_e\right)\ln 2}\!\triangleq\! \widetilde{R}^{\textup{WD}}_S.
	\end{align}
	\hrulefill\vspace*{0pt}
\end{figure*}
where $P^{(j)}_A$ and $P^{(j)}_S$ denote the local points in the $j$-th iteration. To this end, problem  \eqref{Problem_NCT_Power} is further transformed to
\begin{subequations}\label{Problem_NCT_Power_relaxed}
	\begin{align}
		&\label{P_NCT_P_R_C0}\  \max_{P_w,P_v} \widetilde{R}^{\textup{WD}}_S \\
		&\label{P_NCT_P_R_C1}\quad \  {\rm s.t.}\quad \eqref{P_NCT_P_C1}
	\end{align}
\end{subequations}
Now, the form of this problem can be solved directly by existing solver, such as CVX \cite{cvx}.

By solving problems \eqref{Problem_NCT_relaxed1} and \eqref{Problem_NCT_relaxed2}, the pinching beamforming for each waveguide can be directly obtained. In addition, the convergence of the SCA method has been well established in the literature \cite{SCA}. Therefore, by iteratively solving problem \eqref{Problem_NCT_Power_relaxed} based on the obtained pinching beamforming, a stable and reliable solution can be achieved. The overall procedure is outlined in \textbf{Algorithm \ref{alg:B}}. The computational complexity of the algorithm is given by $\mathcal{O}\left(JM^3+2N\right)$, where $J$ represents the number of SCA iterations and $M$ is the number of power allocation variables, which is set to $M=2$ in this paper.
\subsection{Proposed AO Algorithm for the WM Architecture}
Considering that each waveguide in the WM architecture simultaneously transmits both the legitimate signal and AN, the solution developed for the WD architecture is not directly applicable. To tackle this challenge, we propose an AO algorithm tailored for the WM architecture. Specifically, the original problem \eqref{Problem_DW} is decomposed into two subproblems: pinching beamforming and baseband beamforming. These subproblems are then solved in an iterative manner using the PSO method and the SCA method, respectively.
\begin{algorithm}[!t]\label{Algorithm 2}
	\caption{Proposed Two-Stage Algorithm For Solving problem \eqref{Problem_DW} under the WD Architecture.}
	\label{alg:B}
	\begin{algorithmic}[1]
		\STATE \textbf{Stage 1: Pinching Beamforming Design}\\
		\STATE {Solve problems \eqref{Problem_NCT_relaxed1} and \eqref{Problem_NCT_relaxed2} to determine the positions of all PAs on waveguide 1 and waveguide 2 using the \textbf{Algorithm \ref{alg:A}}.} \\
		\STATE \textbf{Stage 2: Baseband Power Allocation Optimization}
		\STATE Initialize the variables $\{P^{0}_A,P^{0}_S\}$, convergence criterion $\epsilon_0$. Let $j=1$.
		\REPEAT 
		\STATE {Given the $\{\mathbf{x}_1,\mathbf{x}_2\}$ and $\{P^{(j-1)}_w,P^{(j-1)}_v\}$, solve probblem \eqref{Problem_NCT_Power_relaxed}} iteratively using SCA method to determine the $P^{(j)}_w$ and $P^{(j)}_v$.\\
		\STATE Update $j \leftarrow j+1$.\\
		\UNTIL the iteration yield is below $\varepsilon_0$.\\
		\STATE {Output} the optimal solutions $\mathbf{x}_1^*$, $\mathbf{x}_2^*$, $P^{*}_w$ and $P^{*}_v$.
	\end{algorithmic}
\end{algorithm}

\emph{1) Pinching Beamforming Optimization:} With fixed baseband beamforming vectors $\mathbf{w}$ and $\mathbf{v}$, the original problem \eqref{Problem_DW}  is transformed to
\begin{subequations}\label{Problem_CT_PA}
	\begin{align}
		&\label{P_C_P_C0}\ \max_{\{\mathbf{x}_1,\mathbf{x}_2\}} \quad R_S\\ 
		&\label{P_C_P_C1}\quad {\rm s.t.} \ \eqref{P_DW_C2},\eqref{P_DW_C3}.
	\end{align}
\end{subequations}
Due to the diversity and complexity of the optimization variables, traditional optimization methods struggle to effectively solve this problem. To overcome this challenge, we employ the heuristic PSO algorithm as a suitable alternative. To start with, we define a particle swarm of size $I$ and initialize each particle associated with the optimization variables as follows
\begin{align}
	\mathbf{X}_i^{(0)}=[x^{(0)}_{i,1},\cdots,x^{(0)}_{i,N},x^{(0)}_{i,N+1},\cdots,x^{(0)}_{i,2N}]^T,
\end{align}
where $x_{i,n}$ represents the $x$-coordinate of the $n$-th PA on waveguide 1 in the $i$-th particle, while $x_{i,N+n}$ represents the $x$-coordinate of the $n$-th PA on waveguide 2 in the $i$-th particle. To satisfy constraint \eqref{P_DW_C2}, we let all  $x^{(0)}_{i,n}$ take values between $\left[-\frac{L}{2},\frac{L}{2}\right]$. Correspondingly, we define the initial velocity of these particle swarms as follows
\begin{align}
	\mathbf{U}_i^{(0)}=[u^{(0)}_{i,n},\cdots,u^{(0)}_{i,N},u^{(0)}_{i,N+1},\cdots,u^{(0)}_{i,2N}]^T.
\end{align}

Following the principles of the PSO algorithm \cite{PSO}, each particle updates its position in each iteration based on its own historically best position and the globally best position of the entire swarm. Therefore, we might let $\mathbf{X}_{i,p}$ and $\mathbf{X}_{g}$ denote the personal best position of the $i$th particle and the global best position of the entire swarm, respectively. Based on this, the iterative update equation can be expressed as follows
\begin{align}\label{velocity}
	\mathbf{U}^{(l+1)}_{i}\!\!=\alpha\mathbf{U}^{(l)}_i\!+\!c_1\beta_1(\mathbf{X}_{i,p}\!-\!\mathbf{X}^{(l)}_i)\!+\!c_2\beta_2(\mathbf{X}_{g}\!-\!\mathbf{X}^{(l)}_i),
\end{align}
\begin{align}\label{position}
	\mathbf{X}^{(l+1)}_i=\mathbf{X}^{(l)}_i+\mathbf{U}^{(l+1)}_i,
\end{align}
where $l$ denotes the iteration index, $\alpha$ represents the inertia weight, which controls the influence of the particle's previous velocity on its current velocity, helping balance exploration and exploitation. Notably, a larger $\beta$ value enhances global search capability but weakens personal search capability, while a smaller $\alpha$ value has the opposite effect. Therefore, to achieve better performance, we employ a linearly decreasing $\alpha$ over iterations, varying within the range $\left[\alpha_{\min},\alpha_{\max}\right]$, with an update step of $\left(\alpha_{\max}-\alpha_{\min}\right)/L_{\max}$. Besides, the coefficients $c_1$ and $c_2$ serve as learning factors that determine the relative impact of the personal and global best positions on the particle's velocity update. Correspondingly, the random variables $\beta_1$ and $\beta_2$ are uniformly distributed between 0 and 1, introducing stochasticity to enhance the diversity of the search process and prevent premature convergence. It is important to emphasize that after each position update, a boundary check must be performed. If any element in the particle exceeds the range $\left[-\frac{L}{2},\frac{L}{2}\right]$, it should be adjusted to the nearest boundary value to ensure that constraint \eqref{P_DW_C2} is always satisfied.

In each iteration, the fitness function of each particle is evaluated based on the unconstrained problem \eqref{Problem_CT_PA}. Specifically, given the PAs' positions $\mathbf{X}_i$ according to the $i$-th particle, the SR can be calculated from \eqref{SR} and denoted by $\mathcal{R}_S\left(\mathbf{X}_i\right)$. Furthermore, in order to ensure constraint \eqref{P_DW_C3}, we introduce a penalty factor $\xi>0$ to the fitness function and update it as follows 
\begin{align} \label{fitness}
	\mathcal{F}(\mathbf{X}_i)=\mathcal{R}(\mathbf{X}_i)-\xi\mathcal{P}(\mathbf{X}_i).
\end{align}
where $\mathcal{P}(\mathbf{X}_i)$ is a penalty function that counts the number of instances in which the current PA position violates the constraint \eqref{P_DW_C3}, which is given by
\begin{align}
	\mathcal{P}(\mathbf{X}_i)&=\sum_{n=2}^{N} \mathbb{I}(\mathbf{x}_{i,n}-\mathbf{x}_{i,n-1}<\Delta) \nonumber \\
	&+\sum_{n=N+2}^{2N} \mathbb{I}(\mathbf{x}_{i,n}-\mathbf{x}_{i,n-1}<\Delta),
\end{align}
where $\mathbb{I}$ represents an indicator function that equals 1 if the condition inside the parentheses holds and 0 otherwise. It is important to note that to enforce the distance constraint between all pinching antennas, the penalty factor 
$\xi$ should typically be assigned a sufficiently large value. With the fitness evaluation conducted on each particle, their local and global best positions are improved until convergence. 

\emph{2) Baseband Beamforming Optimization:} With the given PA position vectors $\{\mathbf{x}_1,\mathbf{x}_2\}$, the original problem degenerates into a conventional PLS beamforming problem. Let $\mathbf{W}\triangleq\mathbf{w}\mathbf{w}^H \in \mathbb{C}^{2\times 2}$ and $\mathbf{V}\triangleq\mathbf{v}\mathbf{v}^{H} \in \mathbb{C}^{2\times 2}$ denote the legitimate signal beamforming matrix and AN beamforming matrix, respectively. In this case, this subproblem is formulated as a semi-positive definite program (SDP) problem \cite{Huang_SDP}:
\begin{subequations}\label{Problem_CT_Beamforming}
	\begin{align}
		&\label{P_CT_B_C0}\  \max_{\mathbf{W},\mathbf{V}} \quad R^{\textup{WM}}_b-R^{\textup{WM}}_e \\ 
		&\label{P_CT_B_C1}\quad \  {\rm s.t.}\quad \mathrm{Tr}\left(\mathbf{W}\right)+\mathrm{Tr}\left(\mathbf{V}\right) \leq P_{\max}, \\
		&\label{P_CT_B_C2}\quad \quad \quad \quad \mathbf{W}, \mathbf{V} \succeq 0, \\
		&\label{P_CT_B_C3}\quad \quad \quad \quad \mathrm{Rank}\left(\mathbf{W}\right)=1, \\
		&\label{P_CT_B_C4}\quad \quad \quad \quad \mathrm{Rank}\left(\mathbf{V}\right)=1.
	\end{align}
\end{subequations}
Due to the non-convexity of the objective function and rank-one constraints, this problem is a challenge to solve in its current form. In fact, leveraging the theoretical proofs from \cite[\textbf{Remark 1}]{Rank-one}, we can directly solve a relaxed version of the problem by omitting the rank-one constraints \eqref{P_CT_B_C3} and \eqref{P_CT_B_C4}, ensuring that the obtained solution remains consistent with the original problem. For the non-convex objective function, we can employ the SCA method to relax this objective function. According to the first order Taylor expansion, we have the convex lower boundary of the objective function as \eqref{R_CT_low}, as shown at the top of the next page, 
\begin{figure*}[t]
	\normalsize
		\setlength{\belowdisplayskip}{-0.3pt}
	\begin{align}\label{R_CT_low}
		&\log_2\left(1+\!\frac{\mathrm{Tr}\left(\mathbf{h}^H_b\mathbf{W}\mathbf{h}_b\right)}{\mathrm{Tr}\left(\mathbf{h}^H_b\mathbf{V}\mathbf{h}_b\right)+\sigma_b^2}\right)-\log_2\left(1+\frac{\mathrm{Tr}\left(\mathbf{h}^H_e\mathbf{W}\mathbf{h}_e\right)}{\mathrm{Tr}\left(\mathbf{h}^H_e\mathbf{V}\mathbf{h}_e\right)\!+\!\sigma_e^2}\!\right) \nonumber \\
	&=\log_2\!\left(\!\mathrm{Tr}\!\left(\mathbf{h}^H_b\!\mathbf{W}\mathbf{h}_b\!\right)\!\!+\!\!\mathrm{Tr}\!\left(\mathbf{h}^H_b\!\mathbf{V}\mathbf{h}_b\!\right)\!\!+\!\!\sigma^2_b\right)\!\!+\!\log_2\!\left(\!\mathrm{Tr}\!\left(\mathbf{h}^H_e\!\mathbf{V}\mathbf{h}_e\!\right)\!\!+\!\!\sigma^2_e\right)\!\!-\!\log_2\!\left(\mathrm{Tr}\!\left(\!\mathbf{h}^H_b\!\mathbf{V}\mathbf{h}_b\!\right)\!\!+\!\sigma^2_e\right)\!\!-\!\log_2\!\left(\!\mathrm{Tr}\!\left(\mathbf{h}^H_e\!\mathbf{W}\mathbf{h}_e\!\right)\!\!+\!\!\mathrm{Tr}\!\left(\mathbf{h}^H_e\!\mathbf{V}\mathbf{h}_e\!\right)\!\!+\!\!\sigma^2_e\right) \nonumber \\
	&\geq \log_2\left(\mathrm{Tr}\left(\mathbf{h}^H_b\mathbf{W}\mathbf{h}_b\right)\!+\!\mathrm{Tr}\left(\mathbf{h}^H_e\mathbf{V}\mathbf{h}_e\right)\!+\!\sigma^2_b\right)\!+\!\log_2\left(\mathrm{Tr}\left(\mathbf{h}^H_e\mathbf{V}\mathbf{h}_e\right)\!+\!\sigma^2_e\right)\!-\!\log_2\left(\mathrm{Tr}\left(\mathbf{h}^H_e\mathbf{W}^{(j)}\mathbf{h}_e\right)\!+\!\mathrm{Tr}\left(\mathbf{h}^H_e\mathbf{V}^{(j)}\mathbf{h}_e\right)\!+\!\sigma^2_e\right) \nonumber \\
	&-\log_2\left(\!\mathrm{Tr}\left(\mathbf{h}^H_b\mathbf{V}^{(j)}\mathbf{h}_b\!\right)\!+\!\sigma^2_e\right)\!-\!\frac{\mathrm{Tr}\left(\mathbf{h}^H_b\left(\mathbf{V}\!-\!\mathbf{V}^{(j)}\right)\mathbf{h}_b\right)}{\left(\!\mathrm{Tr}\left(\mathbf{h}^H_b\mathbf{V}^{(j)}\mathbf{h}_b\right)\!+\!\sigma^2_e\right)\ln 2}\!-\!\frac{\mathrm{Tr}\left(\mathbf{h}^H_e(\mathbf{W}\!-\!\mathbf{W}^{(j)})\mathbf{h}_e\right)\!+\!\mathrm{Tr}\left(\mathbf{h}^H_e(\mathbf{V}\!-\!\mathbf{V}^{(j)})\mathbf{h}_e\right) }{\left(\mathrm{Tr}\left(\mathbf{h}^H_e\mathbf{W}^{(j)}\mathbf{h}_e\right)\!+\!\mathrm{Tr}\left(\mathbf{h}^H_e\mathbf{V}^{(j)}\mathbf{h}_e\right)+\sigma^2_e\right)\ln 2}\!\triangleq\! \widetilde{R}^{\textup{WM}}_S.
	\end{align}
	\hrulefill\vspace*{0pt}
\end{figure*}
where $\mathbf{W}^{(j)}$ and $\mathbf{V}^{(j)}$ denote the local point in the $j$-th iteration. At this point, problem \eqref{Problem_CT_Beamforming_relaxed} is relaxed to
\begin{subequations}\label{Problem_CT_Beamforming_relaxed}
	\begin{align}
		&\label{P_CT_B_R_C0}\  \max_{\mathbf{W},\mathbf{V}} \widetilde{R}^{\textup{WM}}_S\\ 
		&\label{P_CT_B_R_C1}\quad \  {\rm s.t.}\quad \eqref{P_CT_B_C1}, \eqref{P_CT_B_C2}.
	\end{align}
\end{subequations}
This is a standard convex problem, which can be solved using CVX \cite{cvx}.
\begin{algorithm}[!t]\label{Algorithm 3}
	\caption{Proposed AO Algorithm For Solving Problem \eqref{Problem_DW} under the WM Architecture.}
	\label{alg:C}
	\begin{algorithmic}[1]
		\STATE {Initilaize the particle swarm with $\{\mathbf{X}^{(0)}_i$,$\mathbf{U}^{(0)}_i\}$, initilize the beamforming matrices $\{\mathbf{W}^{(0)},\mathbf{V}^{(0)}\}$. Introduce auxiliary variables $\{\widehat{\mathbf{X}},\widehat{\mathbf{W}},\widehat{\mathbf{V}}\}$}. Set convergence criterion $\varepsilon_1$.\\
		\STATE Let $\widehat{\mathbf{W}}=\mathbf{W}^{(0)}$, $\widehat{\mathbf{V}}=\mathbf{V}^{(0)}$, $\widehat{\mathbf{X}}=\mathbf{X}_g$.
		\REPEAT 
		\STATE Given $\{\widehat{\mathbf{W}},\widehat{\mathbf{V}}\}$, solve problem \eqref{Problem_CT_PA} using PSO method:
		\STATE Initialize the particle swarm with $\{\mathbf{X}^{(0)}_i$,$\mathbf{U}^{(0)}_i\}$, evaluate the fitess function $\mathcal{F}$ value to obtain the local best position $\mathbf{X}_{i,p}$ for each particle, let the global best position $\mathbf{X}_g=\widehat{\mathbf{X}}$, set the iteration index $l=1$.
		\REPEAT 
		\STATE Update the inertia with $\alpha=\alpha_{\max}-\frac{(\alpha_{\max}-\alpha_{\min})l}{L_{\max}}$.
		\REPEAT 
		\STATE Update the velocity and position of particle $i$ according to \eqref{velocity} and \eqref{position}, respectively.\\
		\STATE Evaluate the fitess function $\mathcal{F}$ value for particle $i$ according to \eqref{fitness}.
		\STATE \textbf{if} $\mathcal{F}(\mathbf{X}^{(l)}_{i})> \mathcal{F}(\mathbf{X}_{i,p})$ \textbf{then}
		\STATE  \quad Update $\mathbf{X}_{i,p} \leftarrow \mathbf{X}^{(l)}_{i}$.
		\STATE  \textbf{end if}
		\STATE \textbf{if} $\mathcal{F}(\mathbf{X}^{(l)}_{i})> \mathcal{F}(\mathbf{X}_{g})$ \textbf{then}
		\STATE  \quad Update $\mathbf{X}_{g} \leftarrow \mathbf{X}^{(l)}_{i}$.
		\STATE  \textbf{end if}
		\STATE $i \leftarrow i+1$.
		\UNTIL $i>I$.
		\STATE $t \leftarrow t+1$.
		\UNTIL $t>T$.\\
		\STATE Let $\widehat{\mathbf{X}}=\mathbf{X}_g$.
		\STATE Given $\widehat{\mathbf{X}}$, solve problem \eqref{Problem_CT_Beamforming_relaxed} using SCA method:
		\STATE Let $\mathbf{W}^{0}=\widehat{\mathbf{W}}$, $\mathbf{V}^{0}=\widehat{\mathbf{V}}$, set convergence criterion $\epsilon_1$ and the iteration index $j=1$.
		\REPEAT 
		\STATE Given {$\mathbf{W}^{(j-1)},\mathbf{V}^{(j-1)}$, solve problem \eqref{Problem_CT_Beamforming_relaxed} iteratively to determine $\mathbf{W}^{(j)},\mathbf{V}^{(j)}$}.\\
		\STATE Update $j \leftarrow j+1$.\\
		\UNTIL the iteration yield is below $\varepsilon_2$.\\
		\STATE Let $\widehat{\mathbf{W}}=\mathbf{W}^{(j)}$, $\widehat{\mathbf{V}}=\mathbf{V}^{(j)}$.
		\UNTIL the iteration yield is below $\varepsilon_1$.\\
		\STATE {Output} the optimal solutions $\mathbf{x}_1^*$, $\mathbf{x}_2^*$, $\mathbf{W}^*$ and $\mathbf{V}^*$.
	\end{algorithmic}
\end{algorithm}

The steps for solving problem \eqref{Problem_DW} are clearly outlined in \textbf{Algorithm \ref{alg:C}}. In the PSO method, each iteration begins by initializing the global optimal position of the particle swarm using the optimal result from the previous iteration. Likewise, each SCA iteration is performed based on the last obtained optimal result. As a result, the alternating optimization of problems \eqref{Problem_CT_PA} and \eqref{Problem_CT_Beamforming_relaxed}, whose objective functions are non-decreasing and bounded by the channel capacity, ensures convergence to a stable point after multiple iterations. On the one hand, the computational complexity of the PSO method is given by $\mathcal{O}\left(L_{\max}IM^2\right)$, where $L_{\max}$ denotes the maximum number of iterations, $I$ is the number of particles, and $M$ denotes the number of waveguides with $M=2$. On the other hand, the relaxed SDP problem \eqref{Problem_CT_Beamforming_relaxed} can be solved using the interior-point method \cite{SDR}, whose computational complexity is given by $\mathcal{O}\left(JM^{3.5}\right)$, where $J$ represents the number of SCA iterations. Therefore, the overall computational complexity is given by $\mathcal{O}\left(\frac{1}{\epsilon_1}\left(L_{\max}IM^2+JM^{3.5}\right)\right)$.

\section{Numerical Results}
In this section, numerical results are provided to validate the effectiveness of PASS-enabled secure communications in both single-waveguide and multiple-waveguide scenarios.
\subsection{Simulation Setup}
Unless specified otherwise, the simulation parameters used in this paper are as follows: the height of all waveguides is set to $H=2$ m, the communication carrier frequency is set to $f_c=28$ GHz, the maximum transmit power at the BS is set to $P_{\max}=1$ mW, the noise power is set to $\sigma^2_b=\sigma^2_e=-90$ dBm, and the minimum spacing between PAs to avoid coupling effect is set to half a wavelength, i.e., $\Delta=\frac{\lambda_2}{2}$. Besides, the effective refractive index is set to $n_{eff}=1.4$. Furthermore, in the PSO method, we set $\alpha_{\max}=0.9$, $\alpha_{\max}=0.1$, $c_1=c_2=1.5$, $L_{\max}=300$, and $\xi=100$. While for the convergence of the proposed algorithms, we set $\varepsilon_0=\varepsilon_1=\varepsilon_2=10^{-3}$.
\subsection{Secrecy Performance of the Single-Waveguide Scenario}
In this subsection, we analyze the secrecy performance of the single-waveguide scenario. For clarity in simulation demonstrations, we refer to our proposed algorithm as the \textbf{PAST}. Additionally, we consider three baselines:
\begin{itemize}
	\item\textbf{PSO:} In this scheme, the PSO algorithm is employed to optimize the pinching beamforming in the single-waveguide scenario. A particle swarm comprising $I$ particles is initialized, where each particle has $N$ dimensions corresponding to the positions of the $N$ PAs. The SR is adopted as the fitness function, and particle positions are iteratively updated according to \textbf{Algorithm \ref{alg:C}} to determine the optimal PA configuration.
	\item\textbf{Random Position:} In this scheme, the positions of all PAs are randomly generated without any optimization. The final result is obtained by averaging the SR over 500 independent realizations.
	\item\textbf{Conventional Antenna:} In this scheme, a BS equipped with the same number of antennas as the PAs is deployed, but only a single RF chain is activated and fixed at location $\left[-L/2, 0, H\right]^T$. To maximize the SR, analog beamforming is employed at the BS.
\end{itemize}

\begin{figure}[]
	\setlength{\abovecaptionskip}{0cm}   
	\setlength{\belowcaptionskip}{-0.2cm}   
	\setlength{\textfloatsep}{7pt}
	\centering
	\subfigure[Even case.]{\label{even}
		\includegraphics[width= 3.4in]{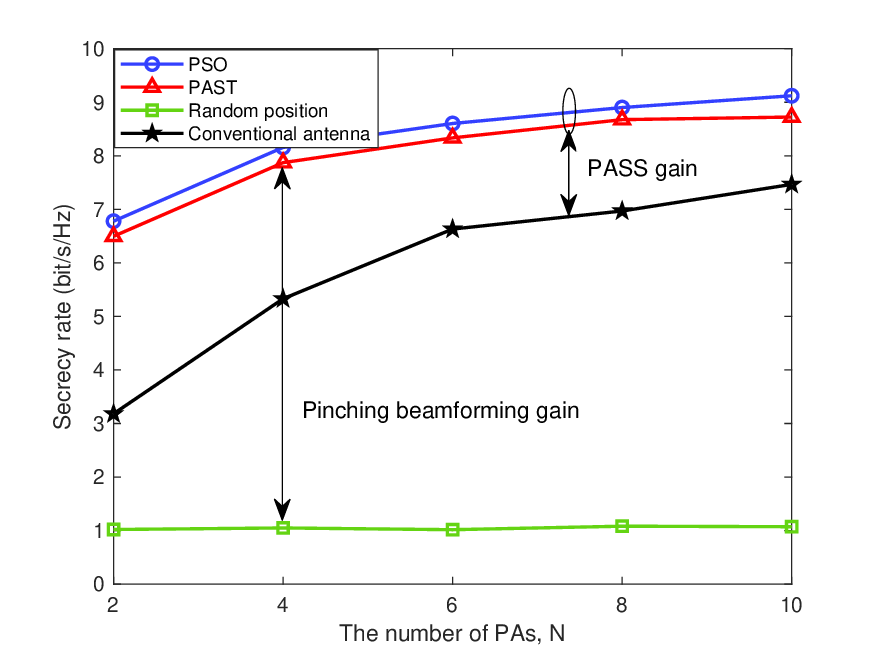}}
	\subfigure[Odd case.]{\label{odd}
		\includegraphics[width= 3.4in]{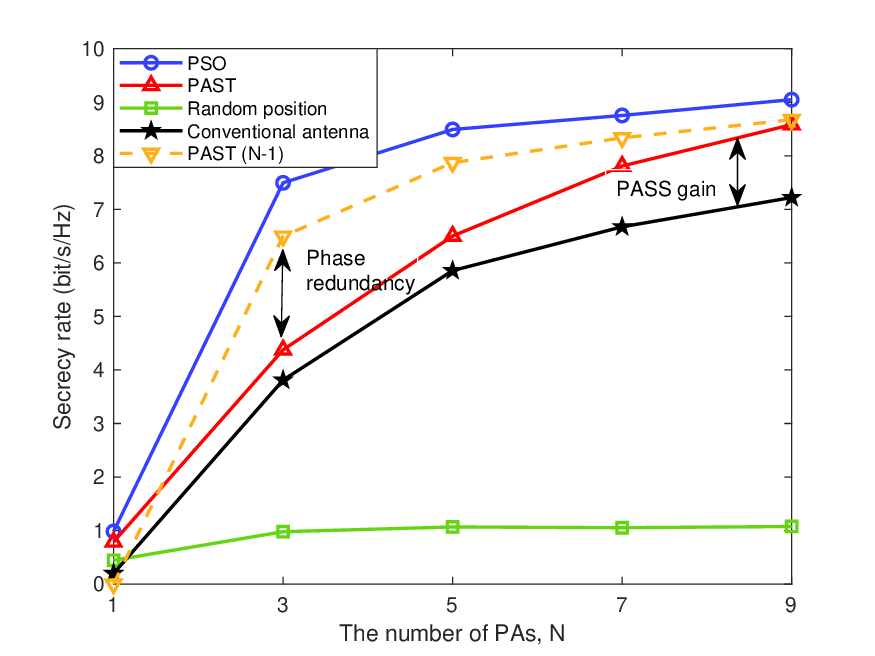}}
	\caption{\textcolor{black}{SR versus the number of PAs, $N$.} } 
	\label{res_SW_PAs}
\end{figure}
Fig. \ref{res_SW_PAs} plots the achievable SR versus the number of PAs, $N$, for $L=5$ m. Particularly, Fig. \ref{even} presents the case where $N$ is an even number. It is observed that, except for the random position scheme, all methods show improved the SR performance as $N$ increeses. This behavior arises because, in the single-waveguide scenario, the SR primarily depends on the channel quality disparity between Bob and Eve. If Eve's channel quality surpasses Bob's, it can completely intercept the communication, reducing Bob's SR to zero. Consequently, in the random position scenario, even with an increased number of PAs, their random spatial distribution leads to stochastic channel conditions, which fail to provide significant SR enhancement for Bob. In contrast, the other three schemes leverage the increasing $N$ to achieve greater design degrees of freedom (DoFs), enhancing the channel quality differences, thereby improving the SR. This underscores the necessity of optimizing PA positions. Additionally, it can be also observed that the PASS consistently outperforms the conventional antenna systems. This is due to its ability to mitigate the large-scale path loss through the flexible pinching beamforming during transmission. Furthermore, regarding the performance comparison between our proposed PAST and PSO methods, although our proposed algorithm's performance is slightly lower than that of the PSO method, which is mainly caused by phase errors during the PA fine-tuning phase, it offers significantly lower complexity by avoiding extensive searches. This highlights the effectiveness of our proposed PAST algorithm.

Fig. \ref{odd} further depicts the SR performance when $N$ is an odd number. The key difference is that at $N=3$, our proposed PAST algorithm shows a significant performance gap compared to the PSO method, but this gap decreases as $N$ increases. This is because, when $N$ is odd, the phase cancellation process at Eve in the second stage of our method only leaves one PA’s signal uncanceled. As a result, Eve still receives substantial signal information, reducing the SR. However, as $N$ increases, the transmission power of each PA decreases, weakening this effect and allowing the PAST algorithm to approach the performance of the PSO method. To better illustrate this behavior, we also present the case of $N-1$ (i.e., an even number of PAs) for comparison. As expected, it can be observed that the phase redundancy at $N=3$ makes its performance weaker than at $N=2$. This insight provides practical guidance for deployment: when $N$ is small and odd, it may be beneficial to deactivate one PA to employ our proposed PAST algorithm for the pinching beamforming design.

\begin{figure}
	\setlength{\abovecaptionskip}{0cm}   
	\setlength{\belowcaptionskip}{-0.2cm}   
	\setlength{\textfloatsep}{7pt}
	\centering
	\includegraphics[width=3.4in]{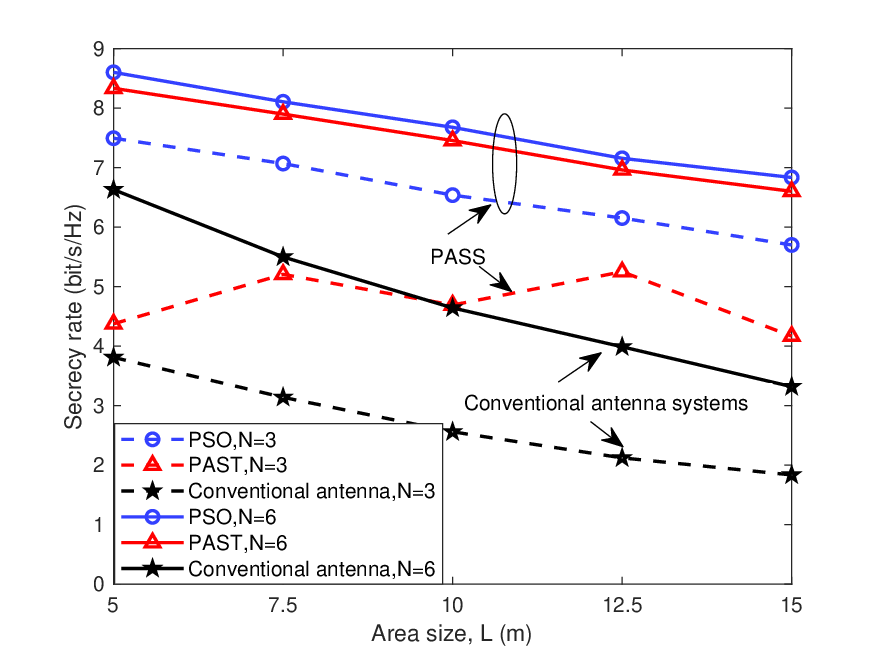}
	\caption{SR versus the area size, $L$}
	\label{res_SW_area}
\end{figure}
In Fig. \ref{res_SW_area}, we investigate the SR versus the area size, $L$. We consider the cases of $N=3$ and $N=6$. It can be found that, expect for the case of $N=3$ under the PAST, the secrecy performance of the other schemes decreases as $L$ increases. This is because an increase in $L$ spreads the users over a wider area, causing the signals to travel longer distances from the BS, resulting in a higher path loss. Moreover, the PASS handles this challenge more effectively than conventional antenna systems, as their flexible mobility allows them to minimize long-distance path loss and serve users through shorter-distance LoS channels. For the case of $N=3$ under the PAST, since it cannot effectively suppress Eve's eavesdropping and Eve's position changes randomly, this causes the curve to fluctuate.

\subsection{Secrecy Performance of the Multiple-Waveguide Scenario}
In this subsection, we analyze the secrecy performance of the multiple-waveguide scenario. Here,  we assume that the two waveguides are symmetrically and uniformly distributed about the $y$-axis with a separation of $D/2$. If not specified, we set $D = 0.5$ m. For performance comparison, we consider the following four baselines: 
\begin{itemize}
	\item  \textbf{WM Without AN}: In this scheme, no AN is introduced at the baseband, and both waveguides are solely used to transmit the legitimate signals.
	\item \textbf{Hybrid beamforming (HB)}: In this scheme, a BS equipped with two RF chains, each connected to $N$ conventional antennas, is deployed at the position $\left[-L/2,0,H\right]^T$. Secure transmission is achieved through the use of AN and hybrid beamforming techniques.
	\item \textbf{Fully-digital beamforming (FDB)}: In this scheme, a BS equipped with $2N$ RF chains and $2N$ antennas (one per RF chain) is deployed at the position $\left[-L/2,0,H\right]^T$. AN is introduced, and fully-digital beamforming is performed to enhance PLS performance.
	\item \textbf{FDB without AN}: This scheme is based on the FDB setup but excludes the use of AN.
\end{itemize}

\begin{figure}
	\setlength{\abovecaptionskip}{0cm}   
	\setlength{\belowcaptionskip}{-0.24cm}   
	\setlength{\textfloatsep}{7pt}
	\centering
	\includegraphics[width=3.4in]{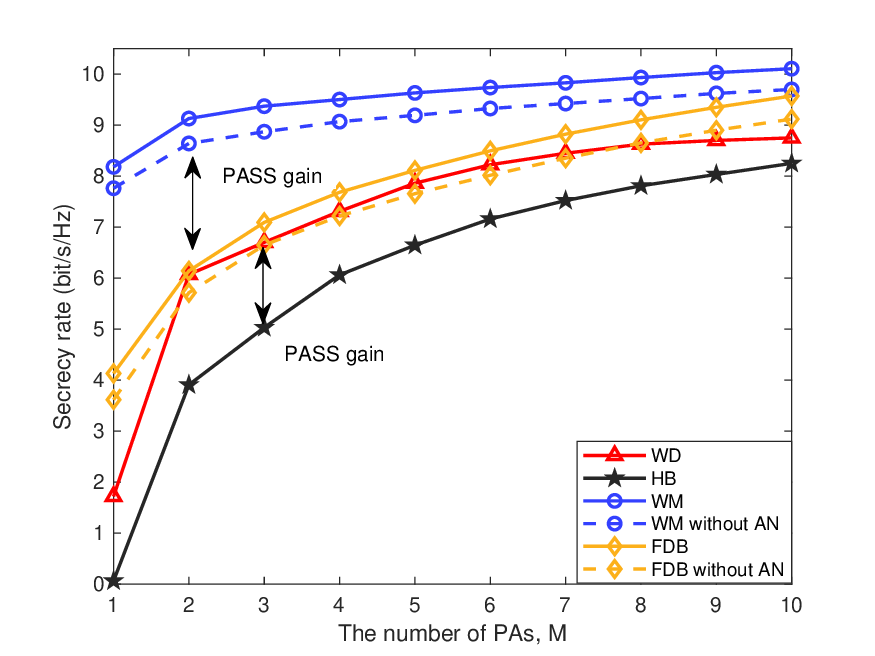}
	\caption{SR versus the number of PAs, $N$}
	\label{res_DW_PAs}
\end{figure}
In Fig. \ref{res_DW_PAs}, we study the SR versus the number of PAs, $N$, where $L=5$ m. It can be seen that all schemes exhibit performance improvements as $N$ increases. This is because an increase in $N$ provides more DoFs to improve the secrecy performance. Compared to conventional antenna systems, PASS can achieve significant gains due to their advantages in coping with the large-scale path loss. Among all schemes, WM achieves the highest SR. This is expected since WM not only leverages the flexible pinching beamforming of PAs to optimize system performance but also fully exploits the baseband’s signal processing capabilities, leading to more precise legitimate signal and AN beams. While WD maintains a simpler structure and achieves performance comparable to FDB and consistently outperforms HB, demonstrating that it offers a good balance between performance and system complexity. As $N$ increases, conventional antenna systems reduce the gap with PASS by utilizing spatial DoFs, but this comes with significantly higher hardware and power costs. In contrast, PASS offers similar scalability at much lower cost, underscoring its economic practicality. Additionally, schemes with AN consistently outperform their AN-free counterparts, which also validates the importance of AN in improving the secrecy performance. 

\begin{figure}
	\setlength{\abovecaptionskip}{0cm}   
	\setlength{\belowcaptionskip}{-0.15cm}   
	\setlength{\textfloatsep}{7pt}
	\centering
	\includegraphics[width=3.4in]{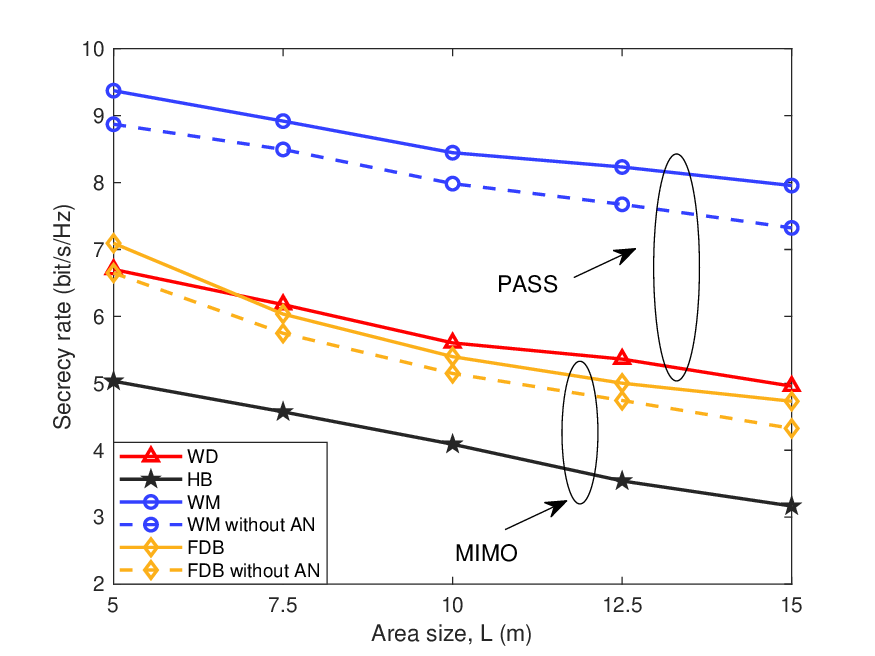}
	\caption{SR versus the area size, $L$}
	\label{res_DW_area}
\end{figure}
Fig. \ref{res_DW_area} plots the SR versus the area size, $L$. We set $N=3$. It can be seen that the performance of all the schemes decreases as $L$ expands. This is because the expansion of the user distribution area in general leads to a general increase in the communication distance. In this scenario, PASS is more capable to handle this issue, leading to less performance degradation compared to conventional antenna systems.

\begin{figure}
	\setlength{\abovecaptionskip}{0cm}   
	\setlength{\belowcaptionskip}{-0.15cm}   
	\setlength{\textfloatsep}{7pt}
	\centering
	\includegraphics[width=3.4in]{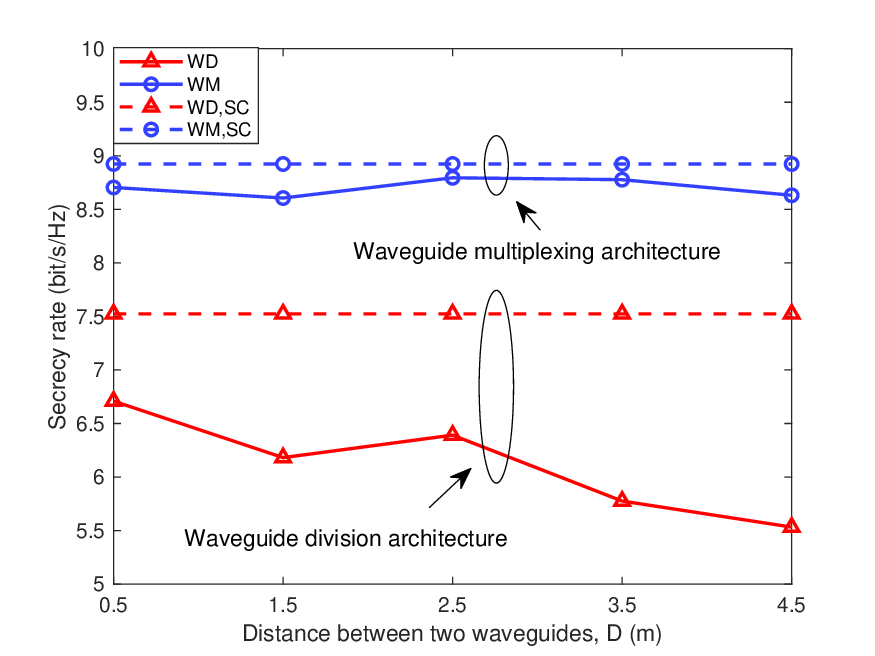}
	\caption{SR versus the waveguide spacing, $D$}
	\label{res_DW_spacing}
\end{figure}
In Fig. \ref{res_DW_spacing}, we further compare the SR achieved by WD and WM with respect to waveguide spacing $D$ while keeping $N=3$ and $L=5$ m. Moreover, we also provide two special cases, namely \textbf{WD, SC} and \textbf{WM, SC}, where two waveguides are deployed directly above each user. As observed, the SR performance of WD degrades significantly with the increase of $D$, whereas WM remains relatively stable with only slight fluctuations. The reasons can be explained as follows: in WD,  since each waveguide transmits either the legitimate signal or AN, increasing the spacing weakens both channel conditions to Bob and Eve, reducing signal strength and jamming effectiveness, thereby causing SR fluctuations. In contrast, WM leverages baseband beamforming to jointly optimize the transmission across waveguides, thereby achieving more balanced signal and AN design that is robust to channel variations caused by waveguide spacing. However, regardless of the transmission architectures, deploying waveguides directly above the users achieves higher performance by ensuring better channel conditions. These results highlight the importance of waveguide placement, which will be a crucial consideration for future PASS deployment and optimization.
\section{Conclusions}
This paper proposed a PASS-enabled secure wireless communication framework. To fully harness the potential of this emerging flexible antenna technology for secure wireless communications, both single-waveguide and multiple-waveguide scenarios were analyzed. In each scenario, the secrecy performance maximization problem was formulated. To solve this non-convex problem, a PAST algorithm was proposed for the single-waveguide scenario. The corresponding numerical results verified its effectiveness, especially when the number of PAs is even or large. For the multi-waveguide scenario, a two-stage algorithm was developed for the WD transmission architecture, while an AO algorithm with PSO and SCA methods was designed for the WM transmission architecture. The numerical results showed that WM is superior to WD, while WD provides a more complexity-efficient alternative with competitive performance. These design insights highlight the potential of PASS as a cost-efficient solution for secure wireless communication, paving the way for future research on PA deployment and optimization strategies. Future work may also consider more general scenarios, such as eavesdroppers equipped with multiple antennas, to further validate and extend the proposed framework.

\section*{Appendix: Proof of Lemma 1} \label{Appendix:A}
Here, we first prove that the optimal spacing between neighboring PAs is $\Delta$. As can be observed, $\frac{1}{\sqrt{(x_n-x_b)^2+D^2_b}}$ is monotonically decreasing with respect to distance. Therefore, we can assume that the optimal location of the $n$-th PA has been determined as $x^*_n$ and the $(n-1)$-th PA is position positioned at a distance of $\Delta+\widetilde{\Delta}$ from it, where $\widetilde{\Delta}\geq 0$. In this case, we can obtain the following inequality
\begin{align}
&\frac{1}{\sqrt{(x_n-x_b)^2+D^2_b}}+\frac{1}{\sqrt{(x_n+\Delta+\widetilde{\Delta}-x_b)^2+D^2_b}} \nonumber \\
	&\leq \frac{1}{\sqrt{(x_n-x_b)^2+D^2_b}}+\frac{1}{\sqrt{(x_n+\Delta-x_b)^2+D^2_b}}.
\end{align}
The equality holds if and only if $\widetilde{\Delta}=0$, implying that the objective improves as the spacing decreases. Thus, the optimal PA spacing is $\Delta$.

Next, we prove the solution's symmetry with respect to $x_m$. Due to the monotonicity and symmetry of the path-loss function, any asymmetric solution would increase the average distance to Bob, leading to higher path loss. For $N=1$, the optimal solution is undoubtedly $x^*_1=x_b$. When $N=2$, we assume $x_1=x_b-a$ and $x_2=x_b+b$, where $a,b \geq 0, a+b=\Delta$. Substituting $a$, $b=\Delta-a$ into the original objective function, we can obtain a new function $S(a)$ with respect to $a$
\begin{align}
	S(a)=\frac{1}{a^2+D^2_b}+\frac{1}{(\Delta-a)^2+D^2_b}.
\end{align}
Taking its derivative, we obtain
\begin{align}
	S'(a)=\frac{-2a}{(a^2+D^2_b)^2}+\frac{2(\Delta-a)}{((\Delta-a)^2+D^2_b)^2}.
\end{align}
Continuing the derivation, the second-order derivative of $S(a)$ can be expressed as follows
\begin{align}
	S''(a)=\frac{6a^2-2D^2_b}{(a^2+D^2_b)^3}+\frac{6(\Delta-a)^2-2D^2_b}{((\Delta-a)^2+D^2_b)^3}.
\end{align}
Since $\Delta \ll D_b$, the second derivative $S''(a)$ is negative over the entire interval $[0,\Delta]$, implying that $S'(a)$ is strictly decreasing. Given that $S'(0)>0$, $S'(\Delta)<0$, and $S'(\frac{\Delta}{2})=0$, we conclude that $S(a)$ is strictly increasing on $(0,\frac{\Delta}{2}]$ and strictly decreasing on $[\frac{\Delta}{2},\Delta)$, thus attaining its maximum at $a =\frac{\Delta}{2}$. Therefore, the optimal PA positions are $x^*_1 = x_b-\frac{\Delta}{2}$ and $x^*_2 = x_b + \frac{\Delta}{2}$.

Furthermore, when $N=2k+1$ with $k>1$, we fix the central PA at $x_b$, and determine the remaining $2k$ positions as in the $N=2$ case. For $N=2k$ with $k>2$, the solution is directly obtained by scaling the $N=2$ result. Hence, for any $N$, the optimal PA positions are symmetrically distributed around $x_b$ with step size $\Delta$. This compelets the proof.
\bibliographystyle{IEEEtran} 
\bibliography{pinching.bib}
\end{document}